\documentclass[fleqn,usenatbib]{mnras}

\usepackage{newtxtext,newtxmath}
\usepackage[T1]{fontenc}
\DeclareRobustCommand{\VAN}[3]{#2}
\let\VANthebibliography\thebibliography
\def\thebibliography{\DeclareRobustCommand{\VAN}[3]{##3}\VANthebibliography}

\usepackage{graphicx}	
\usepackage{xcolor}
\usepackage{amsmath}	
\usepackage{newtxtext,newtxmath}   
\DeclareSymbolFont{CMletters}{OML}{cmm}{m}{it}
\DeclareMathSymbol{v}{\mathord}{CMletters}{`v}
\usepackage[capitalise]{cleveref} 
\usepackage{caption} 
\usepackage{subcaption}
\usepackage{tabularx}
\usepackage{multirow}
\usepackage{enumerate}

\usepackage{booktabs}   
\usepackage[inline]{enumitem}
\usepackage{hyperref}
\usepackage{placeins} 
\usepackage{xspace}
\newcommand{\Zwcl}{ZwCl 0634.1+4747\xspace}
\newcommand{\aLBAHBA}{$\alpha_{\rm 53~MHz}^{\rm 144~MHz}$\xspace}
\newcommand{\ainj}{$\alpha_{\rm inj}$\xspace}
\newcommand{\dinj}{$\delta_{\rm inj}$\xspace}
\newcommand{\re}{r_{\rm{e}}}
\newcommand{\reviewfirst}[1]{\textcolor{black}{#1}} 
\newcommand{\reviewsecond}[1]{\textcolor{black}{#1}}

\title[Head-tail radio galaxies in ZwCl0634.1+47474]{Re-energisation of AGN head-tail radio galaxies in the galaxy cluster ZwCl0634.1+47474}

\author[G. Lusetti et al.]{
G. Lusetti,$^{1}$\thanks{E-mail: giulia.lusetti@hs.uni-hamburg.de}
F. de Gasperin,$^{1,2}$
V. Cuciti,$^{1,2}$
M. Br\"uggen,$^{1}$
C. Spinelli,$^{3}$
H. Edler,$^{1}$
G. Brunetti,$^{2}$
\newauthor
R. J. van Weeren,$^{4}$
A. Botteon,$^{2}$
G. Di Gennaro,$^{1}$
R. Cassano,$^{2}$
C. Tasse,$^{6,7}$
T. W. Shimwell$^{4,5}$
\\
$^{1}$Hamburger Sternwarte, Universität Hamburg, Gojenbergsweg 112, 21029 Hamburg, Germany\\
$^{2}$INAF - Istituto di Radioastronomia di Bologna, Via Gobetti 101, 40129 Bologna, Italy \\
$^{3}$Argelander-Institut f\"ur Astronomie Universita\"at Bonn, Auf dem H\"gel 71, 53121 Bonn, Germany\\
$^{4}$Leiden Observatory, Leiden University, PO Box 9513, 2300 RA Leiden, The Netherlands \\
$^{5}$ASTRON, The Netherlands Institute for Radio Astronomy, Oude Hoogeveensedijk 4, 7991 PD, Dwingeloo, The Netherlands \\
$^{6}$GEPI, Observatoire de Paris, CNRS, Université Paris Diderot, 5 place Jules Janssen, 92190 Meudon, France \\
$^{7}$Centre for Radio Astronomy Techniques and Technologies, Department of Physics and Electronics, Rhodes University, Grahamstown 6140, South Africa 
}

\pubyear{2024}

\begin{document}
\label{firstpage}
\pagerange{\pageref{firstpage}--\pageref{lastpage}}
\maketitle

\begin{abstract}
Low-frequency radio observations show an increasing number of radio galaxies located in galaxy clusters that display peculiar morphologies and spectral profiles. This is the result of the dynamical interaction of the galaxy with the surrounding medium.
Studying this phenomenon is key to understanding the evolution of low-energy relativistic particles in the intracluster medium.
We present a multi-frequency study of the three head-tail (HT) radio galaxies and the radio halo in the galaxy cluster ZwCl0634.1+4747. We make use of observations at four frequencies performed with LOFAR LBA (53 MHz), HBA (144 MHz), GMRT (323 MHz) and \reviewfirst{VLA} (1518 MHz) data. The use of extremely low radio frequency observations, such as LOFAR at 53 and 144 MHz, allowed us to detect the extension of the tails up to a distance of $\sim$ 1 Mpc. 
%
We extracted spectral profiles along the tails in order to identify possible departures from a pure ageing model, such as the Jaffe-Perola (JP) model, which only involves synchrotron and inverse-Compton losses. 
We found clear evidence of departures from this simple ageing model, such as surface brightness enhancement and spectral flattening along all of the tails.
This can be interpreted as the consequence of particle re-acceleration along the tails.
Possible explanations for this behaviour include the interaction between a shock and the radio tails or a turbulence-driven re-acceleration mechanism.
We show that the latter scenario is able to reproduce the characteristic features that we observed in our profiles.
\end{abstract}



\section{Introduction}\label{sec:intro}

Galaxy clusters host a large variety of radio emission, generally divided between diffuse radio sources \citep{vanWeeren2019} and radio galaxies \citep{Hardcastle2020_review}. 
These non-thermal emissions are the results of relativistic cosmic-ray electrons (CRe) interacting with the cluster or galactic magnetic field.

%
%
Radio galaxies in clusters of galaxies usually show distorted morphologies \citep{ODeaOwen1985, FerettiGiovannini2008, Garon2019}, because of the dynamical interaction between the radio jets and the surrounding dense ($n_{\rm e}\sim10^{-2}-10^{-4}\,\rm cm^{-3}$), hot ($\rm 10^7 - 10^9\,K$) intra-cluster medium (ICM). 
The first morphological classification by \cite{OwenRudnick1976} is based on the angle between the radio tails and the core of the galaxy \citep[see also][for a review]{Miley1980}. 
Narrow-angle tails \citep[NATs;][]{Venkatesan1994, Feretti1998} have angles between the two jets smaller than $90^\circ$, because of the high ram pressure they experience that leads to the narrow V- or L-shape they assume. 
Wide-angle tails \citep[WATs;][]{Klamer2004, Mao2010} show C-type morphologies with angles between the two tails greater than $90^\circ$ but smaller than $180^\circ$. They are usually thought to experience weaker ram pressure, caused by low velocities relative to the cluster centre and/or lower density of the surrounding ICM.
Among these types, we focus on the head-tail (HT) radio galaxies \citep{RyleWindram1968, HillLongair1971, Miley1973, Sebastian2017, Cuciti2018, Srivastava2020, Lal2020, Muller2021, Botteon2021}, NATs whose distinctive feature is a very elongated structure formed by the radio jets. In HT radio galaxies, both jets are bent in the same direction and thus it is not always possible to resolve them. 
Their radio surface brightness usually peaks close to the host (optical) galaxy and gradually decreases along the tail that can reach up to hundreds of kiloparsecs \citep[e.g.,][]{Srivastava2020, Botteon2021}. Their spectral index becomes steeper, moving from the core towards the tail \citep[e.g.,][]{Cuciti2018, Wilber2018}, while the fractional polarisation increases along the jets \citep{Feretti1998, Muller2021}, reflecting an increase of the degree of ordering of the magnetic field.
However, after the jets are launched into the ICM, relativistic electrons are visible only for tens of Myrs at GHz frequencies because of their energy losses, owing to synchrotron and Inverse-Compton (IC) emission. Thus, these tails left behind by the galaxy moving within the cluster are expected to fade quickly, becoming invisible at high frequencies. Because of the longer cooling times of low-energy CRe, low-frequency observations are crucial for identifying and tracing CRe for much longer.

In fact, with the advent of low-frequency ($<1$ GHz) radio astronomy, tails extending to very long distances have become more common and can be studied in detail \citep{Sebastian2017, Wilber2018, Botteon2021, deGasperin2017MNRAS, Edler2022}. 
In this context, the LOw-Frequency Array \citep[LOFAR;][]{vanHaarlem2013_LOFAR}, being the largest and most sensitive radio-interferometer in the 10-240 MHz regime, has made many recent contributions to the field \citep[e.g.][]{Mandal2020, Pal2023}.
\\
\indent \reviewfirst{Moreover, in some cases the tails show peculiar properties, such as an increase in surface brightness and flattening of the spectral index along the tail \citep[e.g.,][]{Sijbring1998,Parma1999, Giacintucci2007, deGasperin2017MNRAS}.
}
To explain such behaviours far from the host galaxy, the presence of re-energetization processes has been proposed.
In fact, re-acceleration of CRe would be able to power the population of aged electrons initially injected by the AGN, making them visible at larger distances, where they would be expected to disappear owing to radiative losses. 
For example, \cite{deGasperin2017MNRAS} proposed a re-acceleration mechanism invoking turbulence driven into the tail by interactions with the ICM. This would create the so-called gently re-energized tail \citep[GReET;][]{deGasperin2017MNRAS, Edler2022} sources. \cite{Muller2021} suggest the transition from laminar to turbulent flow as the cause of gentle re-acceleration of the electrons along the tail. They also point out that for complex morphologies, more than one AGN cycle, i.e. a change in the injection rate, must be taken into account to explain the entire appearance of the jets. In other cases, adiabatic compression \citep{EnsslinKrishna2001, EnsslinBrueggen2002} would be able to form radio phoenices, revived fossil plasma from old radio lobes.  Radio phoenices usually show an irregular morphology, steep spectral index ($\alpha\sim\, < -1.5 $  with the $S_{\nu}\propto\nu^{\alpha}$ for the synchrotron emission) and curved integrated spectra \citep{Kempner2004, vanWeeren2009, vanWeeren2011, deGasperin2015, Mandal2019, Duchesne2021, Pasini2022}.
A first attempt to systematically study radio phoenices was performed by \citet{Mandal2020}. They found a non-uniform spectral index across the sources, suggesting a possible mix of cosmic-ray populations with different ages, losses, and re-acceleration efficiencies.
Regardless of the type of mechanism that generates them, sources that trace re-energised AGN radio plasma are characterized by very steep spectra. Thus, low-frequency facilities, such as LOFAR, are ideal to study these phenomena.

%
An increasing number of galaxy clusters is also observed to host diffuse Mpc-scale sources, called radio halos \citep[see][for a review]{vanWeeren2019}. Radio halos are steep spectrum ($\alpha < - 1.1$), centrally located sources, whose morphology roughly follows the distribution of the thermal ICM. Turbulent re-acceleration is thought to be the main mechanism responsible for generating radio halos \citep[see e.g.,][]{Brunetti_and_Jones2014}. In this scenario, a population of seed particles are re-accelerated by turbulence generated in the ICM during cluster mergers. 
Thus, the presence of radio halos is connected to the merging history of the systems and the cluster mass, which sets the available energy budget during mergers.
Observational evidences for this scenario are the connection between the dynamical status of the cluster and the presence of radio halos \citep{Buote2001, Cassano2010, Cuciti2015, Cuciti2021} and the correlation between radio halos power and the mass of the host cluster \citep[][]{Cassano2013, Cuciti2021, vanWeeren2021extraction, Duchesne2021,Cuciti2023}.
A key prediction of the turbulent re-acceleration model is that the cutoff in the synchrotron spectrum scales with the energetics of the merger \citep[e.g.,][]{Cassano_Brunetti_2005, Cassano2006}. This means that more massive systems suffering major mergers, host radio halos visible up to $\sim$GHz frequencies, while less massive systems and/or minor mergers produce radio halos with steeper spectra ($\alpha<-1.5$), preferentially detectable at lower frequencies.
These are generally referred to as Ultra-Steep Spectrum radio halos \citep[USSRH;][]{Brunetti2008}.

%
%
\subsection{\Zwcl}\label{sec:ZWCL0634}

\begin{figure}
     \centering
     \includegraphics[width=\columnwidth]{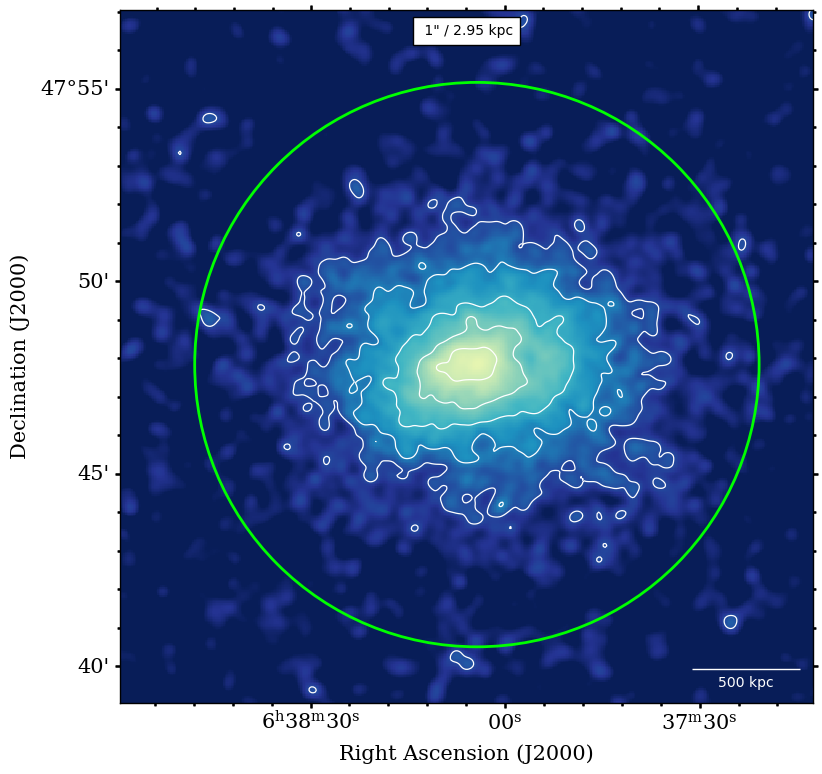}
     \caption{Background-subtracted exposure-corrected XMM image of \Zwcl, in [0.5-2] keV band. The contour levels start at $3\,\cdot$ rms where $\rm rms=5\cdot10^{-6} \,cts\,s^{-1}$. As a size reference, green circle indicates $ R_{\rm 500}=1299$~kpc \citep{Lovisari2017}, namely the radius within which the mean mass over-density of the cluster is 500 times the cosmic critical density at the cluster redshift.}
     \label{fig:XMM_R500}
\end{figure}

\Zwcl is a massive ($M_{500}= (6.65\pm0.33)\times 10^{14}\,\rm M_{\odot}$, \cite{PSZ2_2016}) nearby ($z=0.174$, \cite{Rossetti2017_ref_redshift}) galaxy cluster located at $\rm R.A.(J2000) \, \,06^h 38^m 02^s.5\,,DEC.(J2000)\,\,+47^{\circ}\, 47^{\arcmin}\, 23.8^{\arcsec}$, also known as PSZ1 G167.64+17.63/ PSZ2 G167.67+17.63, CIZA  J0638.1+4747, MCXC J0638.1+4747 or RXC J0638.1+4747.
%
%
Chandra X-ray observations show evidence of a non-relaxed dynamical state, such as the presence of substructures in the X-ray surface brightness distribution \citep{Cuciti2015}. The morphology of the cluster is elongated in the east-west direction (see \cref{fig:XMM_R500}).

\begin{figure}
     \centering
     \includegraphics[width=\columnwidth]{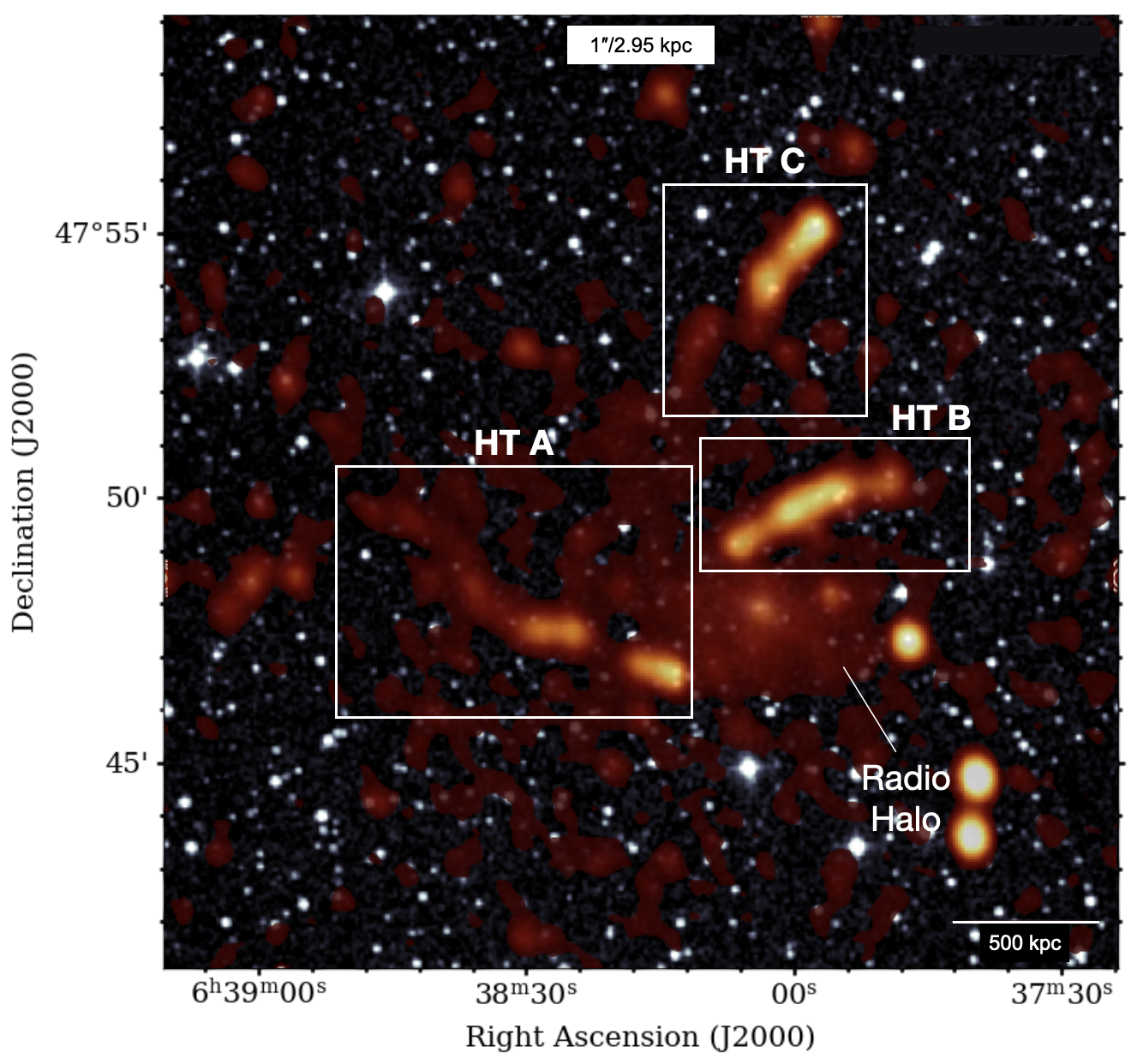}
     \caption{Composite radio-optical image of \Zwcl. LOFAR HBA at 144 MHz superimposed on the DSS image \citep{Paris2014_SDSS-DR10}.}
    \label{fig:final_final_radio+opt_labels_labels}
\end{figure}

\Zwcl hosts a variety of peculiar radio sources. 
%
A radio halo was already detected at 323 and 1518 MHz by \cite{Cuciti2018}. It extends over $\sim$600 kpc in the E-W direction following the morphology of the X-ray emission of the cluster.
%
%
More recently, a radio megahalo has been discovered in this system using LOFAR 144 MHz observations \citep{Cuciti2022Nat}. The megahalo covers the whole cluster volume, with a linear size of 2.8 Mpc and an integrated spectral index between 53 and 144 MHz of $\alpha = -1.62\pm0.25$.
%
In addition, LOFAR observations clearly revealed the presence of three head-tail (HT) radio galaxies in the cluster field,  with $\sim\,0.6-1$ Mpc linear size, labelled HT-A HT-B and HT-C in \cref{fig:final_final_radio+opt_labels_labels}, which are the focus of this paper. The HT-B has been studied at high frequency ($323$ and $1518$ MHz) by \citet{Cuciti2018}. Possible signs of re-energetisation have been found through the analysis of the spectral index and surface brightness along the tail. In that case, the interaction between the tail and a shock has been proposed to explain the emission.

In this paper we present a multi-wavelength radio study of the galaxy cluster \Zwcl, combining data from 53 MHz to 1.5 GHz.
First, we focus our analysis on the three extended HT radio galaxies residing in \Zwcl (\cref{subsec:HT}).
We produce flux density and spectral index profiles along the tails of the HT radio galaxies (\cref{subsec:HT}), comparing them with a standard ageing model (\cref{subsec:JP}). In \cref{subsec:X-ray_properties} we complement the analysis of the cluster with X-ray information using XMM-Newton data.
Additionally, we present a new study of the radio halo down to the very-low frequencies in \cref{subsec:radio_halo}.
In \cref{sec:discussion}, we discuss possible scenarios that involve the interaction of HT radio galaxies with the external ICM.
We summarize our results in \cref{sec:conclusion}.\\


Throughout the paper we adopt a $\Lambda$CDM cosmology with $\mathrm{H_0 = 70 \,km\,s^{-1} Mpc^{-1}}$, $\Omega_\Lambda = 0.7$ and $\Omega_m = 0.3$. Thus, $1''$ corresponds to a physical scale of 2.95 kpc at the redshift of \Zwcl. 
We use the convention $S_{\nu}\propto\nu^{\alpha}$ for the radio synchrotron emission, with $\alpha<0$.


\section{Observations and data reduction}
\label{sec:observation_and_datareduction}

The data used in this work cover a broad range of frequencies, including The LOFAR Low Band Antennas (LBA) at 53 MHz, High Band Antennas (HBA) at 144 MHz, Giant Meterwave Radio Telescope (GMRT) at 323 MHz and Karl G. Jansky Very Large Array (VLA) at 1.5 GHz data. A summary of the observations used is listed in \cref{tab:ObservationRadio}.
GMRT and \reviewfirst{VLA} data of this cluster were already analysed by \cite{Cuciti2018}, while LOFAR LBA and HBA data are extensively described is \cite{Cuciti2022Nat}. Thus, we briefly summarize the data reduction procedure in next sections.

\begin{table*}
    \centering
    \caption{Observational overview: LOFAR LBA, HBA, GMRT and \reviewfirst{VLA} observation. References: (1) \citet{Cuciti2022Nat}, (2) \citet{Cuciti2018} }
    \vspace{-0.1cm}
    \label{tab:ObservationRadio}
    \begin{tabular}{ c c c c c c c }
    \hline
    
    Telescope & Frequency & Bandwidth & Configuration & Time 
    & Observation Date & Reference \\
              & MHz  & MHz                      &     &                &                  &           \\
    \hline
    LOFAR LBA       & 53   & 47 & LBA\_OUTER & 8 h   & 2018 Sep 28   & (1) \\  
    LOFAR HBA       & 144  & 48 & HBA\_DUAL  & 8+8 h 
    &  2019 Oct 27, 2021 Mar 10  (1) \\
    GMRT            & 320  & 32 &     -     & 5  h                      & 2015 Dec 19  & (2)  \\
    \reviewfirst{VLA} L-band     & 1518 & 48 &  D array   &    45 m          & 2015 Oct 19 & (2)  \\
    \reviewfirst{VLA} L-band     & 1518 & 48 &  B array   &     40 m         & 2015 Feb 22 & (2)  \\
    \hline 
    \end{tabular}
    \vspace{-0.3cm}

\end{table*}

\subsection{LOFAR LBA data}\label{subsec:LBA_data}

This cluster was observed at ultra-low frequencies using LOFAR LBA for a total of 8 hours integration time in the frequency range 30 - 77 MHz, using the LBA\_OUTER antenna configuration. Observations were conducted using the multi-beam mode, with one beam continuously pointed at the calibrator and one beam at the target. 
First, following \cite{deGasperin2019}, the calibrator solutions are derived and applied to the target field together with the primary beam correction.
The data of the target field were self-calibrated first for direction-independent, and then for direction-dependent effects \citep{deGasperin2020}.
Finally, the image quality was improved via extraction and self-calibration of a small region around the target, adapting the procedure used for HBA analysis of \cite{vanWeeren2021extraction} for LBA data. 
The final model and solutions from the direction-dependent calibration are used to subtract sources outside a circular region of $23\arcmin$ centered on the target. Then, the dataset is phase-shifted to the centre of this region, corrected for the primary beam in this direction and additional rounds of self-calibration on the target are performed to further improve the image quality.
For more details on the data used, we refer the reader to \cite{Cuciti2022Nat}, where the observations used in this work were first presented.

\subsection{LOFAR HBA data}
\label{subsec:HBA_data}

The LOFAR HBA data used in this work are part of the LOFAR Two-metre Sky Survey \citep[LoTSS;][]{Shimwell2017, Shimwell2019,Shimwell2022}, a low-frequency (120-168 MHz) radio survey of the northern sky. 
Data were processed through the Survey Key Science Project reduction pipeline v2.2 presented in \cite{Tasse2021, Shimwell2022}, which includes direction-independent and direction-dependent calibration (prefactor \citep{deGasperin2019}; killMS \citep{Smirnov2015}; DDFacet \citep{Tasse2018}). 
Finally, the data sets have undergone the extraction procedure generally used for HBA observations to enhance the quality of the image \citep{vanWeeren2021extraction, Botteon2022DR2}. 
For more details, we refer again to \cite{Cuciti2022Nat}, where the observations used in this work were firstly used.

A flux-scale correction factor $f=0.992$ has been applied to the nominal LOFAR HBA map values to align the images with the flux density scale of the LOFAR Two-Metre Sky Survey \citep{Hardcastle2021_HBAflux, Shimwell2022}.

\subsection{GMRT data}\label{subsec:GMRT_data}

\Zwcl has been observed with the GMRT at 320 MHz for a total time on source of $\sim$5 h \citep{Cuciti2018}.
First, amplitude and gain corrections were calculated for the calibrator sources and then transferred to the target. Radio frequency interference (RFI) has been removed automatically using Common Astronomy Software Applications package \citep[\texttt{CASA}][]{McMullin2007_CASA} and the central 220 channels were averaged to 22 to reduce the size of the data-set. Finally, a few steps of phase-only self-calibration were performed to reduce phase variations.
Moreover, similarly to the extraction procedure, bright sources in the field of view were subtracted from the $uv$-data using the peeling method \citep[see][for more details on the data reduction of this dataset]{Cuciti2018}.

\subsection{\reviewfirst{VLA} data}\label{subsec:VLA_data}

\reviewfirst{VLA} L-band observations of \Zwcl include  D and B array configuration with a total time on source of $\sim$45 and 40 mins respectively \citep{Cuciti2018}.
First, the complex gain solutions for the calibrator sources  (and the phase calibrator when present) on the full bandwidth were obtained, applying the bandpass and delay solution beforehand. Then, the calibration tables were applied to the target and an automatic radio frequency interference (RFI) removal was made using \textsc{CASA}. The 48 central channels of each spectral window were averaged to 6 to reduce the size of the data-set. \reviewfirst{All images were corrected for the primary beam attenuation.}

After cycles of calibration and self-calibration on each array configuration, the B and D array observations were combined. Since the two pointings did not coincide perfectly, bright sources far from the phase center are subtracted separately for each spectral window from the $uv$-data before combining the data-set.
The combined B+D array image is used for the HT analysis (\cref{subsec:HT}), while the D array only is used for imaging the extended radio halo (\cref{subsec:radio_halo}).

\subsection{Source subtraction}\label{subsec:sources_subtraction}

\begin{figure*}
    \centering
    \begin{subfigure}[b]{0.49\textwidth}
         \centering
         \includegraphics[width=\textwidth]{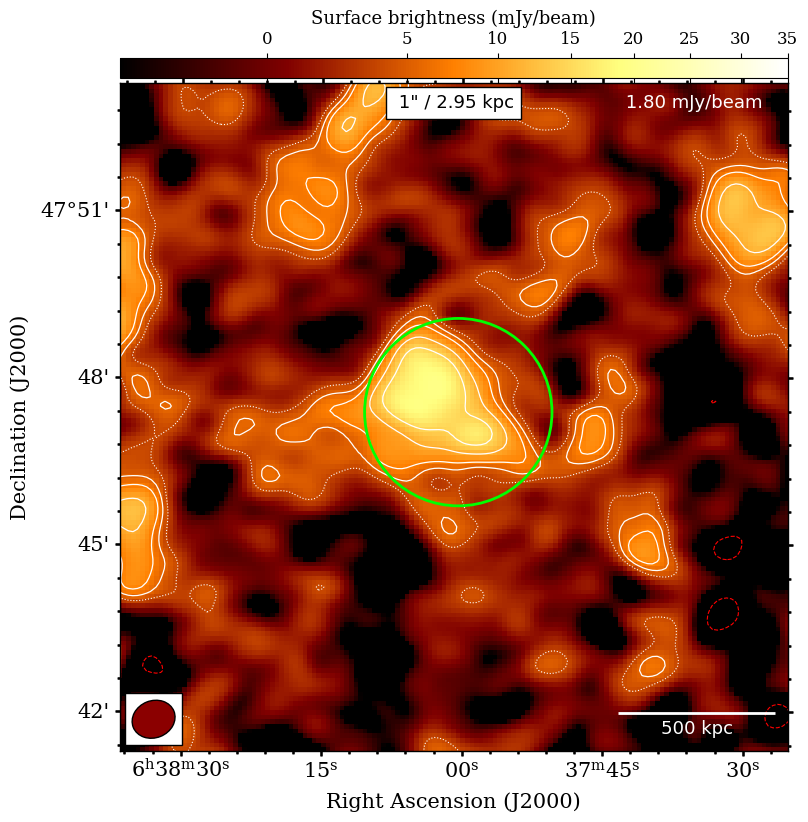}
         \caption{LBA (53 MHz)}
         \label{fig:halo_LBA}
    \end{subfigure}
    \begin{subfigure}[b]{0.49\textwidth}
         \centering
         \includegraphics[width=\textwidth]{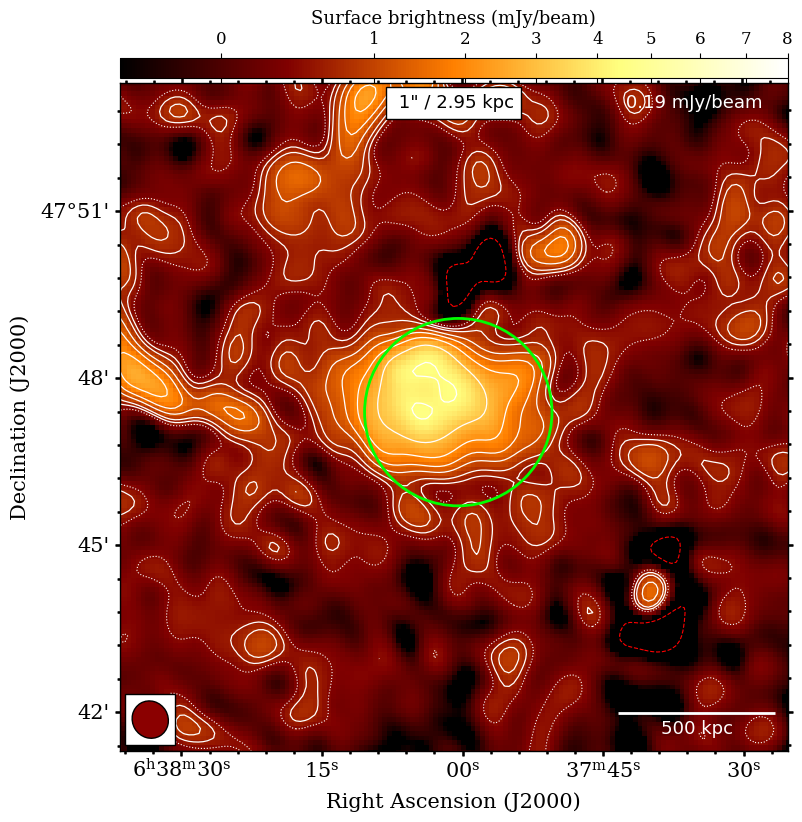}
         \caption{HBA (144 MHz)}
         \label{fig:halo_HBA}
    \end{subfigure}
    \begin{subfigure}[b]{0.49\textwidth}
         \centering
         \includegraphics[width=\textwidth]{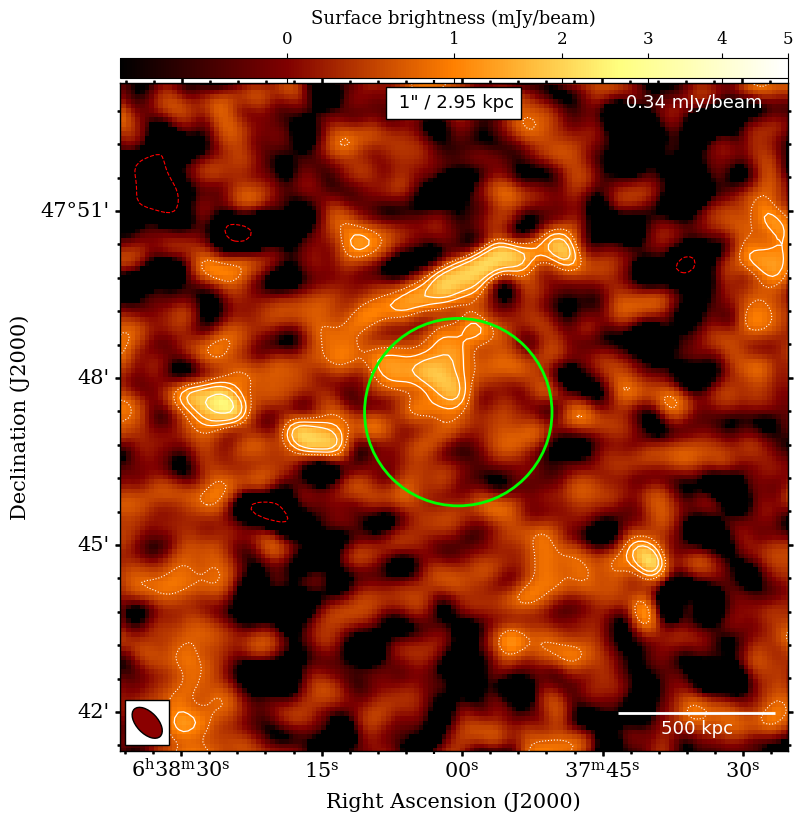}
         \caption{GMRT (323 MHz)}
         \label{fig:halo_GMRT}
    \end{subfigure}
    \begin{subfigure}[b]{0.49\textwidth}
         \centering
         \includegraphics[width=\textwidth]{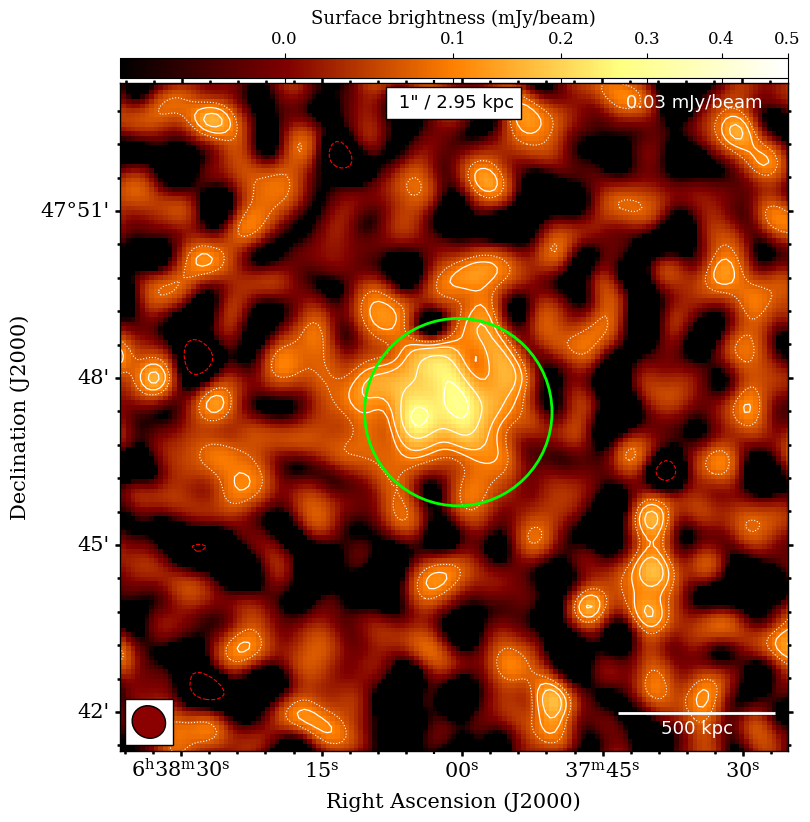}
         \caption{\reviewfirst{VLA} (1518 MHz)}
         \label{fig:halo_VLA}
    \end{subfigure}
    \caption{Low-resolution source-subtracted images of the radio halo hosted by \Zwcl. The beam is shown in the bottom left corner of each image. The contour levels start at $2\sigma_{\rm rms}$ (where $\sigma_{\rm rms}$ is shown in the top right area of the images) and are spaced with a factor of $\sqrt{2}$. The $2\sigma_{\rm rms}$ contours are dotted in white. The $-3\sigma_{\rm rms}$ contours are dashed in red colour. The green circle indicates the region with $R=300$ kpc used to extract the flux densities (\cref{tab:H_summary}). \\
    }
    \label{fig:halos_allfreq}
\end{figure*}

Discrete sources embedded in the extended radio halo, contaminate the diffuse emission and thus, we proceeded to remove them. This allows to properly study the low surface brightness emission from the radio halo.
Previous work \citep{Cuciti2018, Cuciti2022Nat} presented the source subtraction for the low frequency (53 and 144 MHz) and high frequency (323 and 1518 MHz) data set separately. To be consistent through our analysis, we apply the same source subtraction procedure for all four data sets. We describe here briefly the general process.
First, we produced high $(4\arcsec\times5\arcsec)$ resolution images, to identify the compact sources. Then, we selected the corresponding components in the model image and we proceeded to remove them, subtracting these clean components directly from the $uv-$data. 
Finally, we imaged the source-subtracted data sets at low angular resolution (see \cref{fig:halos_allfreq}), maintaining a negative Briggs weighting with robust parameter $-0.25$, using a common inner $uv$-cut of $\mathrm{135\lambdaup}$ (equivalent to an angular scale of $\sim25\arcmin$), which corresponds to the shortest baseline of the \reviewfirst{VLA} data set. 
Moreover, we taper down the long baselines (i.e., using Gaussian taper with a FWHM of $30 \arcsec$) to gain sensitivity towards the diffuse emission. \\
Throughout the paper, the images were created with \textsc{WSClean} software \citep{Offringa2014}. The uncertainty $\Delta_S$ associated with a flux density measurement $S$ is estimated as

\begin{equation}
    \Delta_S = \sqrt{  (\sigma_{\rm c}\cdot S)^2 + N_{\rm beam}\cdot \sigma_{\rm  rms}^2} ,
    \label{eq:flux_error}
\end{equation}
where $N_{\rm beam}=N_{\rm pixel}/A_{\rm beam}$ is the number of independent beams in the source area, and $ \sigma_{\rm c}$ indicates the systematic calibration error on the flux density. Typical values of $\sigma_{\rm c}$ are within 10\% for LOFAR LBA \citep{deGasperin2021}, HBA \citep{Shimwell2022} and GMRT \cite[e.g.][]{Chandra2004}, while 2.5\% is used for \reviewfirst{VLA} \citep{Perley2013}.

\newpage

\subsection{XMM-Newton data}\label{subsec:XMM}

X-ray data for \Zwcl are available in the \textit{XMM-Newton} Science Archive (XSA), OBSID $0692932601$. 
The data reduction of the \textit{XMM-Newton} data is performed using \textsc{HEASoft} version 6.29 \citep{HEAsoft2014} and the Science Analysis Software (\textsc{SAS}\footnote{\href{https://www.cosmos.esa.int/web/xmm-newton/sas}{https://www.cosmos.esa.int/web/xmm-newton/sas}}) version 20.0.0.
Hereafter, we provide a summary of the procedure. For a more extended description of the steps, the reader can refer to \cite{Veronica_2022}.

The first step consists of identifying Good Time Intervals (GTI), filtering out those that are affected by the presence of soft proton flares (SPF) \citep{DeLucaMolendi, KuntzSnowden}. This is achieved by creating a light curve for each of the three instruments of EPIC on board of XMM-Newton, namely MOS1, MOS2 and pn. For all instruments, we also generate count histograms. In the absence of SPF contamination the histogram is well fitted by a Poisson distribution. This allows us to estimate a mean value for the counts and apply a $\pm 3 \sigma$ cut to discard the time intervals affected by flares and, for this reason, falling outside this range.
Additionally, the IN/OUT ratio test \citep{DeLucaMolendi, Leccardi} allows us to further check the quality of our flare filtering procedure by comparing the count rates within the FoV (IN) to those of the unexposed region of the detectors (OUT). An IN/OUT ratio close to 1 for each detector indicates that the contamination by SPF was successfully removed. After the SPF filtering procedure the exposure time is reduced from $\sim 21$ ks to $\sim 13$ ks for the MOS cameras and from $\sim 18$ ks to $\sim 10$ ks for pn.
Note that MOS1-3 and MOS1-6 were not in use at the time of this observation.
Moreover we model and rescale the instrumental background, eventually we subtract it to the observation of \Zwcl and obtain background subtracted and exposure corrected images (\cref{fig:XMM_R500}). The procedure for this step is extensively described in \cite{FWC} and \cite{Migkas2020}. 

Since we are interested in the diffuse X-ray emission associated with the ICM in \Zwcl, we mask the point sources that contaminate the field of view. First, we apply an automated procedure described in \cite{Pacaud2006} which detects point sources and stores them in a catalog. Afterwards, we manually remove any further sources by looking at images in the soft (0.5-2 keV) and hard (2-10 keV) X-ray bands.
The surface brightness analysis, described in Sec.~\ref{subsec:X-ray_properties}, is performed on the masked images.


\section{Results}\label{sec:results}

\begin{table*}
    \centering
    \caption{Images used in this work}
    \vspace{-0.1cm}
    \label{tab:Radio_images}

    \begin{tabular}{ c c c c c c c c }
    \hline
    
    Source & Telescope & Frequency  & Briggs \textbf{weighting} & Taper FWHM & Resolution                  & $\rm \sigma_{\rm rms}$ & Reference \\
           &           & [MHz] &        & [\arcsec]  & [\arcsec$\times$\arcsec]    & [mJy/beam]       &          \\ \hline
    %
    Halo &LBA         & 53    &  -0.25   & 30 & $47\times40$                & $1.80$ &    \cref{fig:halo_LBA}  \\  
         &HBA         & 144   &  -0.25   & 30 & $41\times38$                & $0.19$ &    \cref{fig:halo_HBA}  \\
         &GMRT        & 323   &  0       & 30 & $40\times23$                & $0.34$ &    \cref{fig:halo_GMRT} \\
         &\reviewfirst{VLA}        & 1518  &  -0.25   & 30 & $37\times34$                & $0.03$ &    \cref{fig:halo_VLA}  \\ 
    Head-tails &LBA  & 53     &  -0.25 & 15 & $20\times20$                & $1.22$ &    \cref{fig:tails_LBA}  \\  
               &HBA  & 144    &  -0.25 & 15 & $20\times20$                & $0.12$ &    \cref{fig:tails_HBA} \\
               &GMRT & 323    &  0     & 15 & $20\times20$                & $0.13$ &    \cref{fig:tails_GMRT} \\
               &\reviewfirst{VLA} & 1518   &  -0.25 & 15 & $20\times20$                & $0.04$ &    \cref{fig:tails_VLA} \\ \hline 
    \end{tabular}
\end{table*}


\subsection{Head-tail galaxies}\label{subsec:HT}

\begin{figure*}
    \centering
    \begin{subfigure}[b]{0.49\textwidth}
         \centering
         \includegraphics[width=\textwidth]{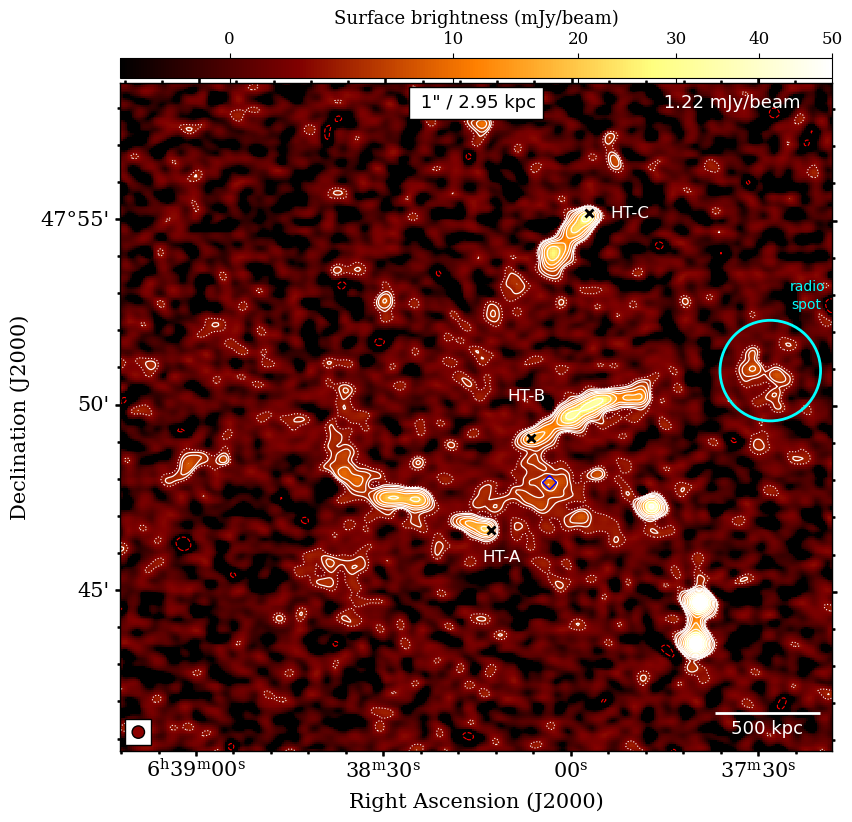}
         \caption{LBA (53 MHz)}
         \label{fig:tails_LBA}
    \end{subfigure}
    \begin{subfigure}[b]{0.49\textwidth}
         \centering
         \includegraphics[width=\textwidth]{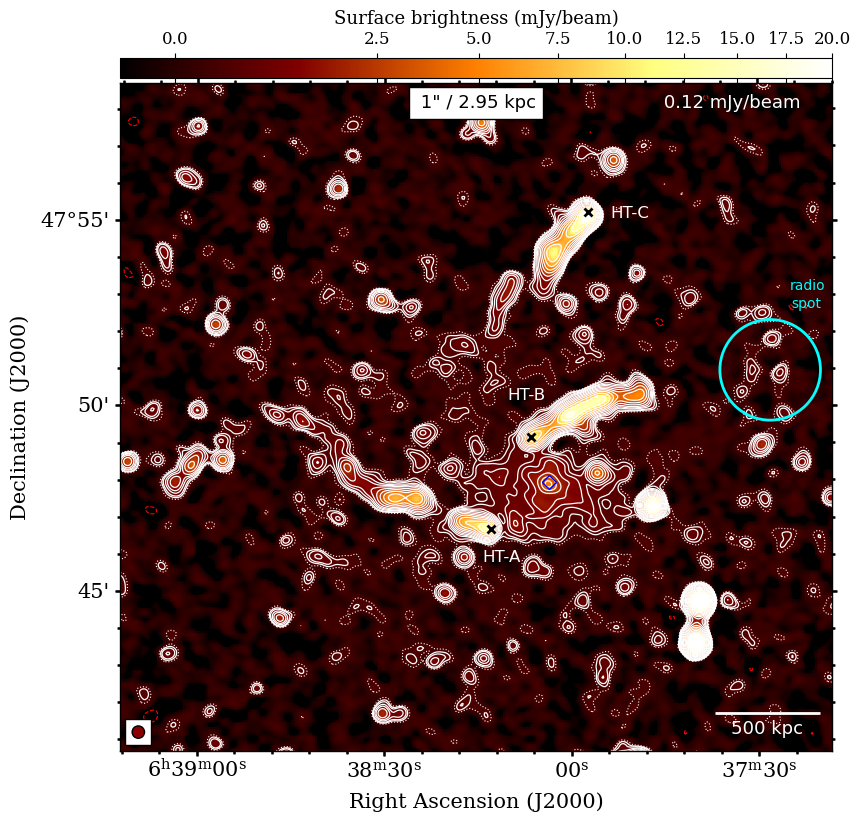}
          \caption{HBA (144 MHz)}
         \label{fig:tails_HBA}
    \end{subfigure}
    \begin{subfigure}[b]{0.49\textwidth}
         \centering
         \includegraphics[width=\textwidth]{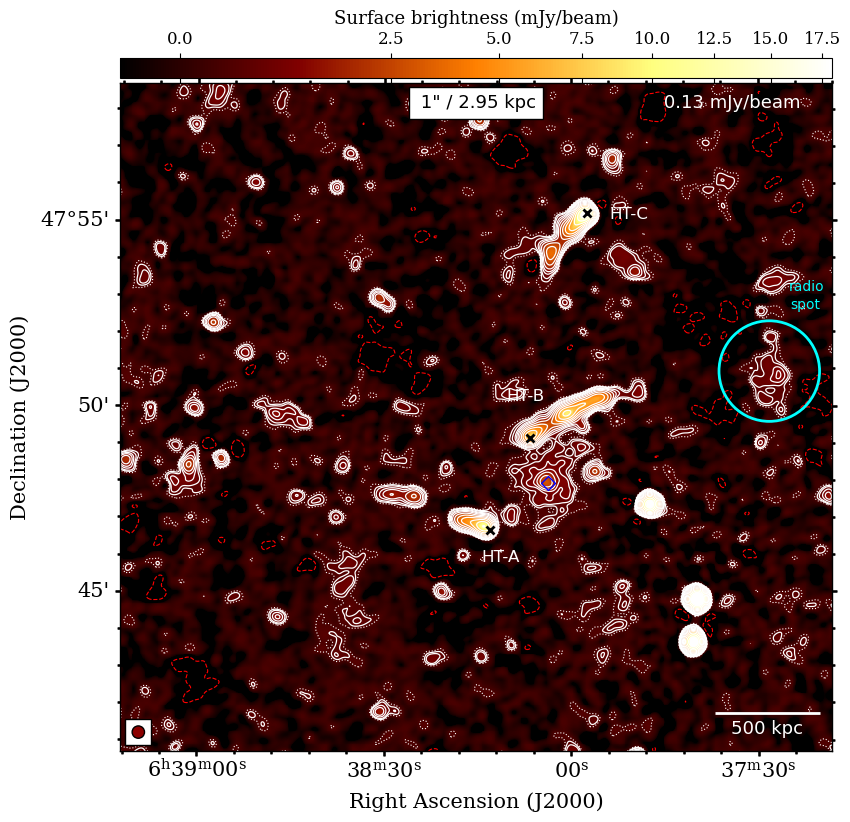}
          \caption{GMRT (323 MHz)}
         \label{fig:tails_GMRT}
    \end{subfigure}
    \begin{subfigure}[b]{0.49\textwidth}
         \centering
         \includegraphics[width=\textwidth]{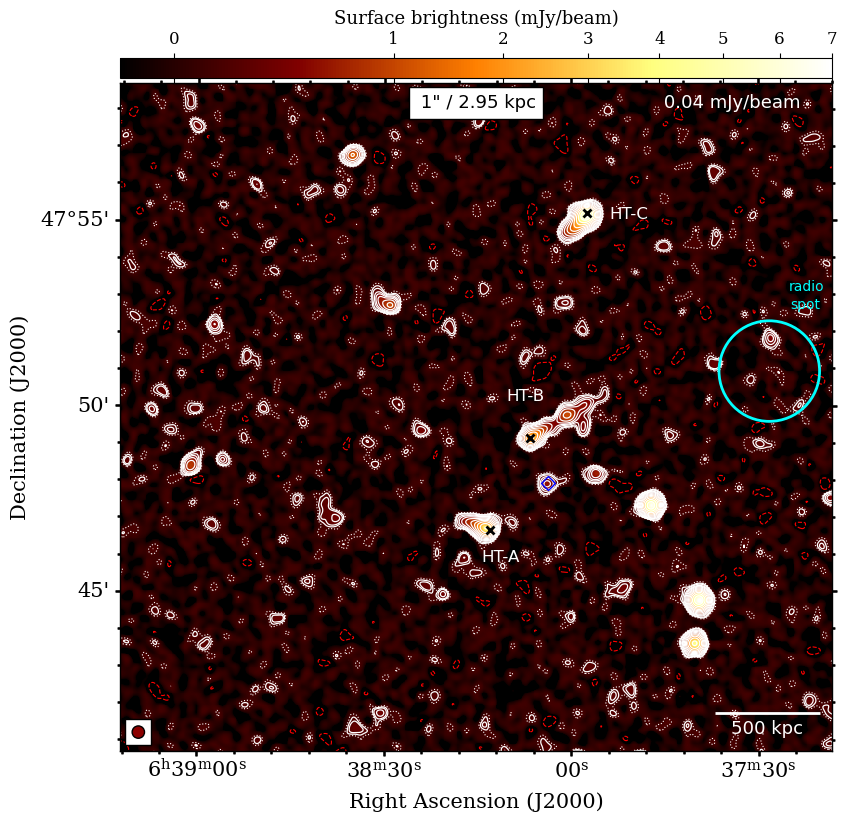}
          \caption{\reviewfirst{VLA} (1518 MHz)}
         \label{fig:tails_VLA}
    \end{subfigure}
    \caption{Images of the head-tail galaxies hosted in \Zwcl at different frequencies. In each image, the contour levels start at $2\sigma_{\rm rms}$, where $\sigma_{\rm rms}$ shown in the top right area of the images, and are spaced with a factor of $\sqrt{2}$. The $2\sigma_{\rm rms}$ are dotted white coloured and $-3\sigma_{\rm rms}$ contours are dashed red coloured. \reviewfirst{The beam is shown in the bottom left corner of each image and has a FWHM of $20\arcsec$.} Black crosses indicate the optical counterparts of the HT radio galaxies (\cref{tab:tails_summary}). The blue diamond is the location of the Brightest Cluster Galaxy (BCG) \citep{Cutri2003}.\\
    }
    \label{fig:tails_allfreq}
\end{figure*}

\begin{figure*}

\centering
     \includegraphics[width=0.9\textwidth]{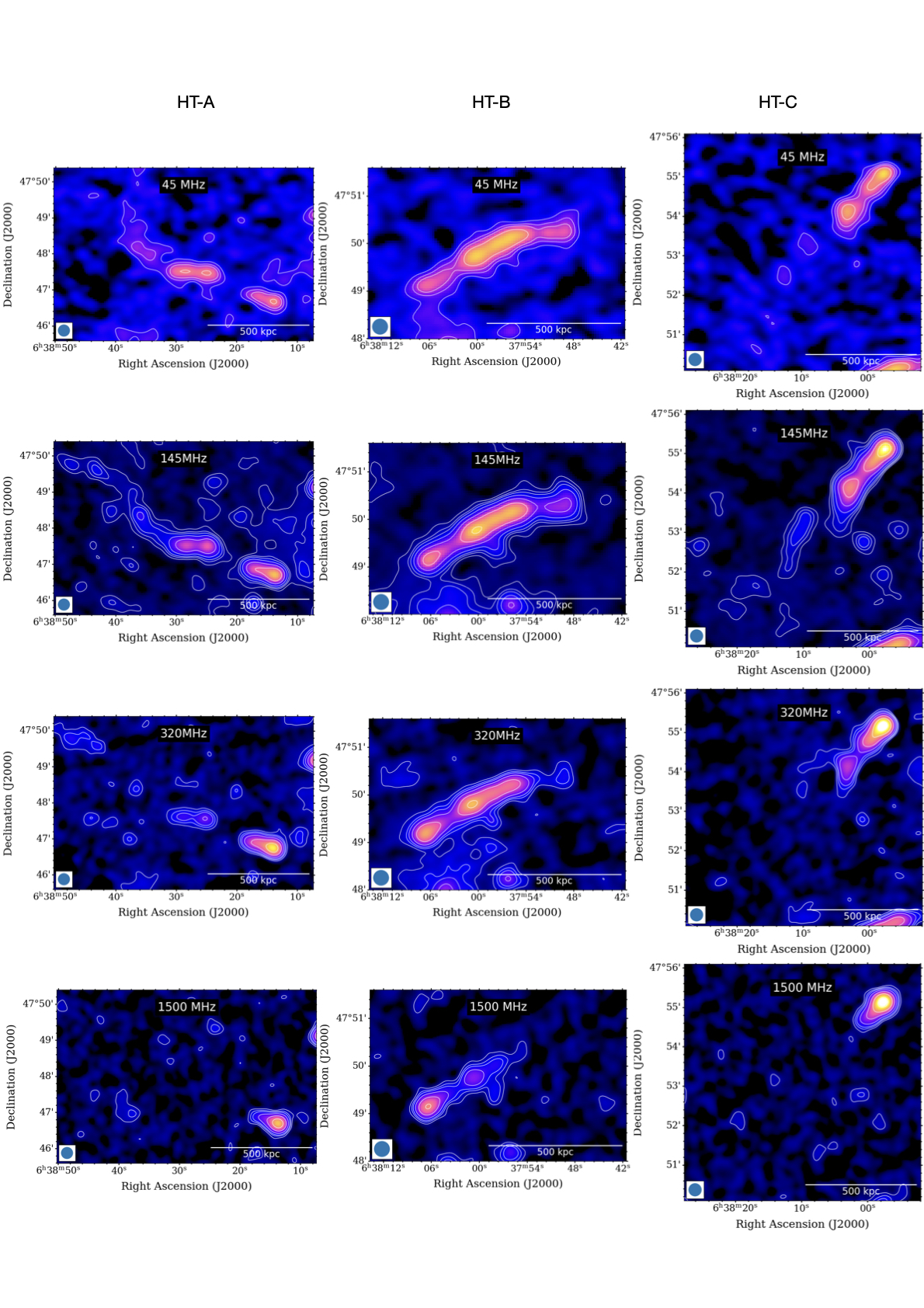}
     \vspace{-1cm}
     \caption{
     \reviewfirst{Zoom-in on the source HT-A \textit{(first column)}, HT-B \textit{(second column)} and HT-C \textit{(third column)} at all frequencies, from 53 MHz \textit{(top)} to 1518 MHz \textit{(bottom)}. For each cut-out, the resolution and $\sigma_{\rm rms}$ are the same as \cref{fig:tails_allfreq}, listed in \cref{tab:Radio_images}.}
     }
     \label{fig:zoomin_tails}
\end{figure*}

The three HT galaxies in \Zwcl are all very extended and bright at lower frequencies (reaching up to $\sim$ 1 Mpc projected size at 144 MHz). They are spread over the cluster volume and show peculiar spectral properties.
We named them HT-A, HT-B and HT-C (see \cref{fig:final_final_radio+opt_labels_labels}). Their properties, including the position and redshift of the optical counterpart, the projected distance from the cluster centre and their maximum extension measured from the $3\sigma_{\rm rms}$ contours of the LOFAR HBA image, are listed in \cref{tab:tails_summary}. 
The cluster-centric distances are projected, and are thus lower limits.
Wide-field, medium-resolution images are shown in \cref{fig:tails_allfreq} while zoomed-in images in \cref{fig:zoomin_tails}. 
The locations of the optical counterparts are shown in \cref{fig:tails_allfreq}. Photometric redshift are taken from the DESI DR9 photometric redshift catalogue \citep{Zhou2021}.

\begin{figure*}
    \centering
     \begin{subfigure}[b]{0.49\textwidth}
         \centering
         \includegraphics[width=\textwidth]{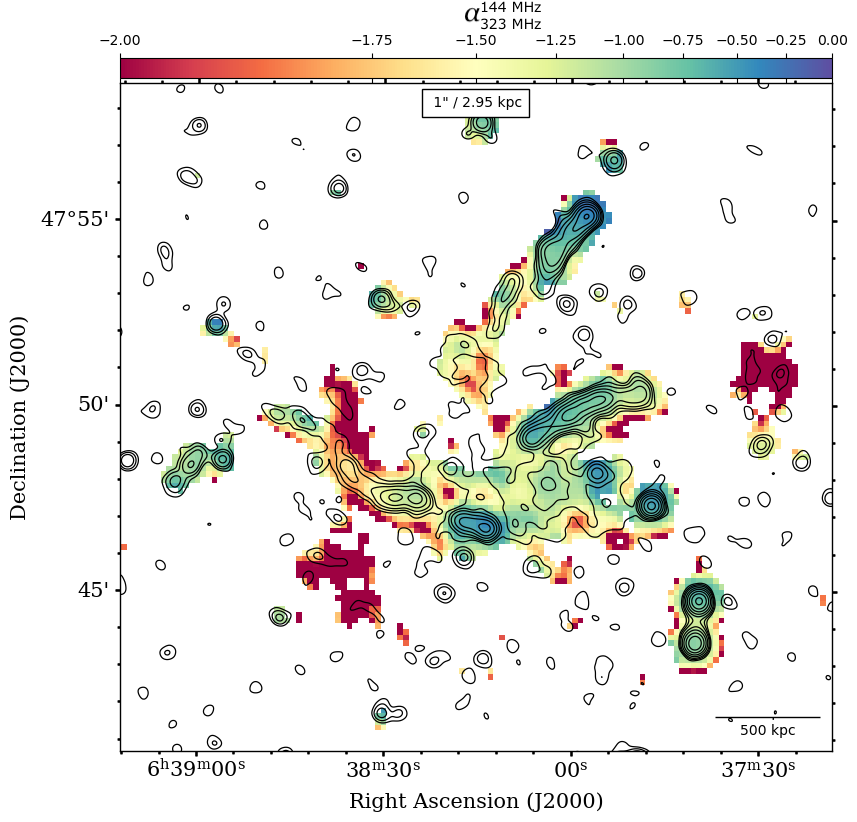}
     \end{subfigure}
     \begin{subfigure}[b]{0.49\textwidth}
         \centering
         \includegraphics[width=\textwidth]{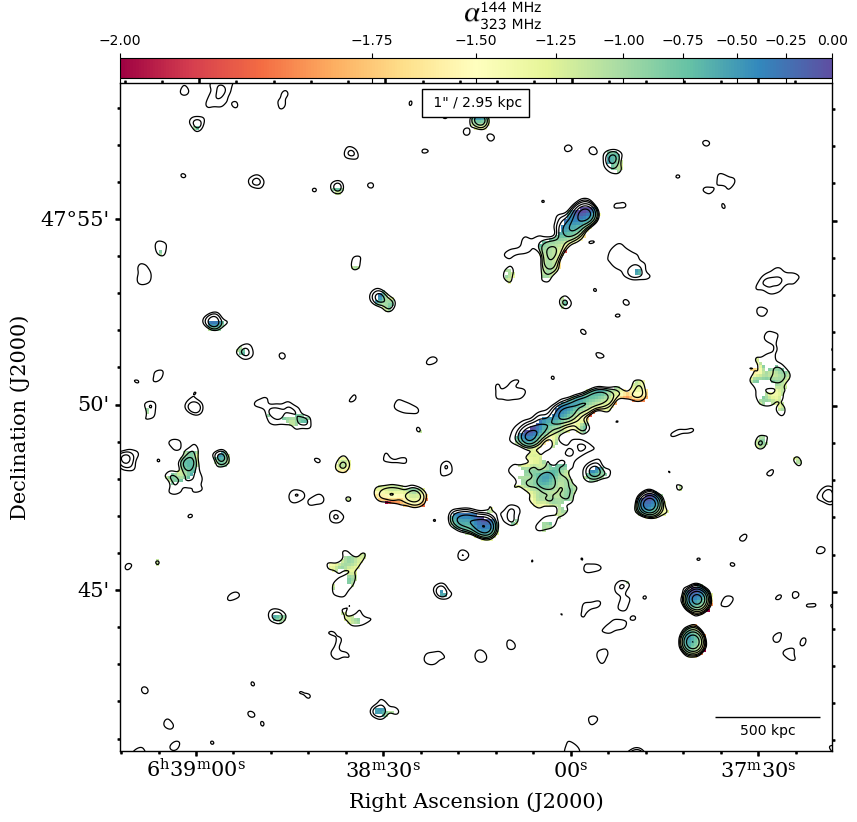}
     \end{subfigure}
    \begin{subfigure}{0.49\textwidth}
         \centering
         \includegraphics[width=\textwidth]{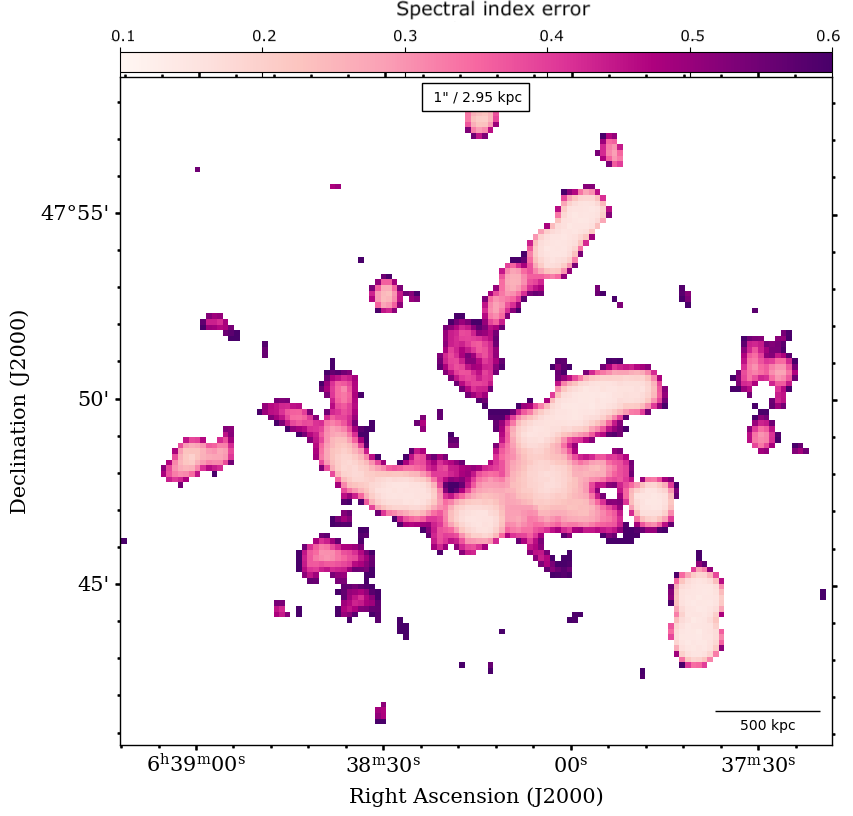}
     \end{subfigure}
    \begin{subfigure}{0.49\textwidth}
         \centering
         \includegraphics[width=\textwidth]{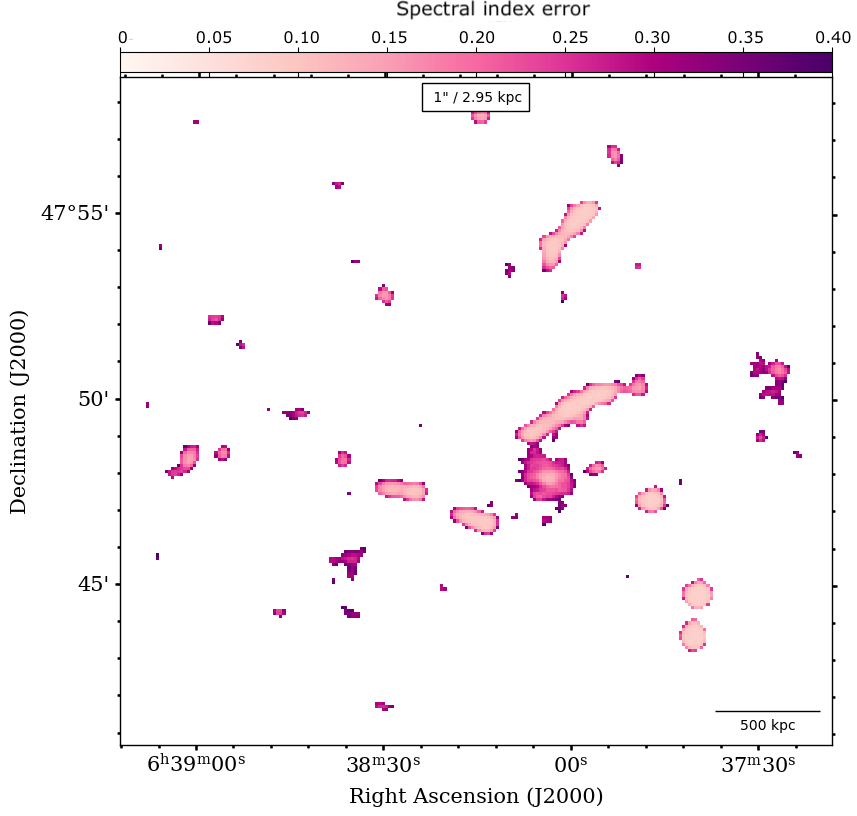}
     \end{subfigure}
     \caption{
     \reviewfirst{\textit{Left panel}: Spectral index map between 53 MHz and 144 MHz at a spatial resolution of $36\arcsec\times36\arcsec$ \textit{(top)} and relative spectral index error $\Delta \alpha_{\rm 144~MHz}^{\rm 53~MHz}$ \textit{(bottom)}. Pixels with surface brightness values below $2\sigma_{\rm rms}$ in the two images were blanked. Black contours start at $3\sigma_{\rm rms}^{\rm 145~MHz}$ and are spaced by a factor of 2. 
     \textit{Right panel}: Spectral index map between 144 MHz and 320 MHz at a spatial resolution of $20\arcsec\times20\arcsec$ \textit{(top)} and relative spectral index error $\Delta \alpha_{\rm 320~MHz}^{\rm 144~MHz}$ \textit{(bottom)}.  Pixels with surface brightness values below $2\sigma_{\rm rms}$ in the two images were blanked. Black contours start at $3\sigma_{\rm rms}^{\rm 320~MHz}$ and are spaced by a factor of 2.}
     }
     \label{fig:spectral_index_map}
\end{figure*}

\begin{figure}
	\includegraphics[width=\columnwidth]{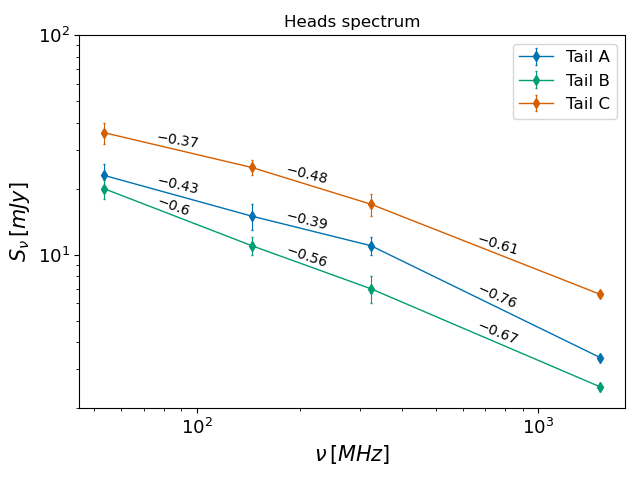}
    \caption{Integrated flux densities of the HTs nuclei, extracted from \cref{fig:tails_allfreq} in regions with $\sim 1.5$ beam area size.}
    \label{fig:heads_tail_spectrum}
\end{figure}

%
Every tail shows a bright head, followed by a lower surface brightness tail, which shows ripples in the radio brightness. 
The presence of multiple radio bumps could be interpreted as other radio-galaxies within the cluster volume rather than a continuation of the tails.
In order to exclude the possibility of a sequence of multiple radio galaxies, we  produced spectral index map between 53~MHz and 144~MHz (\cref{fig:spectral_index_map}, \textit{left}) and 144-320~MHz (\cref{fig:spectral_index_map}, \textit{right}). To this end, we re-imaged LBA and HBA data, using Briggs weighting with Robust parameter $R=-0.25$, common $uv$-range and convolved to the same beam size. In each map, we measured spectral index values for pixels with flux density $\geq 2\sigma_{\rm rms}$ in both frequencies. 
We find that $\alpha_{\tiny{\rm 53~MHz}}^{\tiny{\rm 144~MHz}}$ has flatter values where the heads of the AGN are located (consistent with values \cref{fig:heads_tail_spectrum}) and steepens along the tails. Hence, the bright bumps are unlike other AGN which are expected to have $\alpha\sim -0.5$.
Moreover, no other optical counterparts at the cluster redshift are present in the DESI DR9 photometric redshift catalogue \citep{Zhou2021}.

\begin{table*}
\small
\begin{center}
\renewcommand\arraystretch{1.5}

\caption{\label{tab:tails_summary} Summary of the properties of the head-tail galaxies}
\begin{tabular}{ c c c c c c c c}
\hline
\hline
    Source   & R.A.; DEC & $z_{\rm phot}$&$d$  & $LLS $  &   $\alpha_{\rm inj}$  & $v_{\perp}$\\ 
    
             &            & & \small{kpc} & \small{Mpc}    &       &  \small{km/s}   \\ 
\hline
  HT-A   &$6:38:13.22, +47:46:40.44$& $0.19\pm0.01$ &350 &  1.2     & 0.43  &    $225$       \\ 
  HT-B   &$6:38:06.81,+47:49:05.88 $& $0.16\pm0.01$ &300 &  0.6     & 0.60  &    $274$        \\ 
  HT-C   &$6:37:57.52,+47:55:09.48 $& $0.18\pm0.01$ &870 &  0.8    & 0.37  &    $196$        \\
 \hline
\end{tabular}
\end{center}
\captionsetup{justification=justified}
\caption*{
\begin{enumerate*}[start=1,label={Col. \arabic*:}]
    \item Tail name;
    \item Coordinate of the optical counterpart;
    \item Photometric redshift from \citep{Zhou2021};
    \item Projected distance from the cluster centre;
    \item Largest linear extension above $\rm 3\sigma_{\rm rms}$ at 144 MHz;
    \item injection index $\alpha_{\rm inj}=$\aLBAHBA, shown in \cref{fig:heads_tail_spectrum};
    \item tails' projected velocity;

\end{enumerate*}  }
\end{table*}

\begin{figure*}
    \centering
    \includegraphics[scale=0.42]{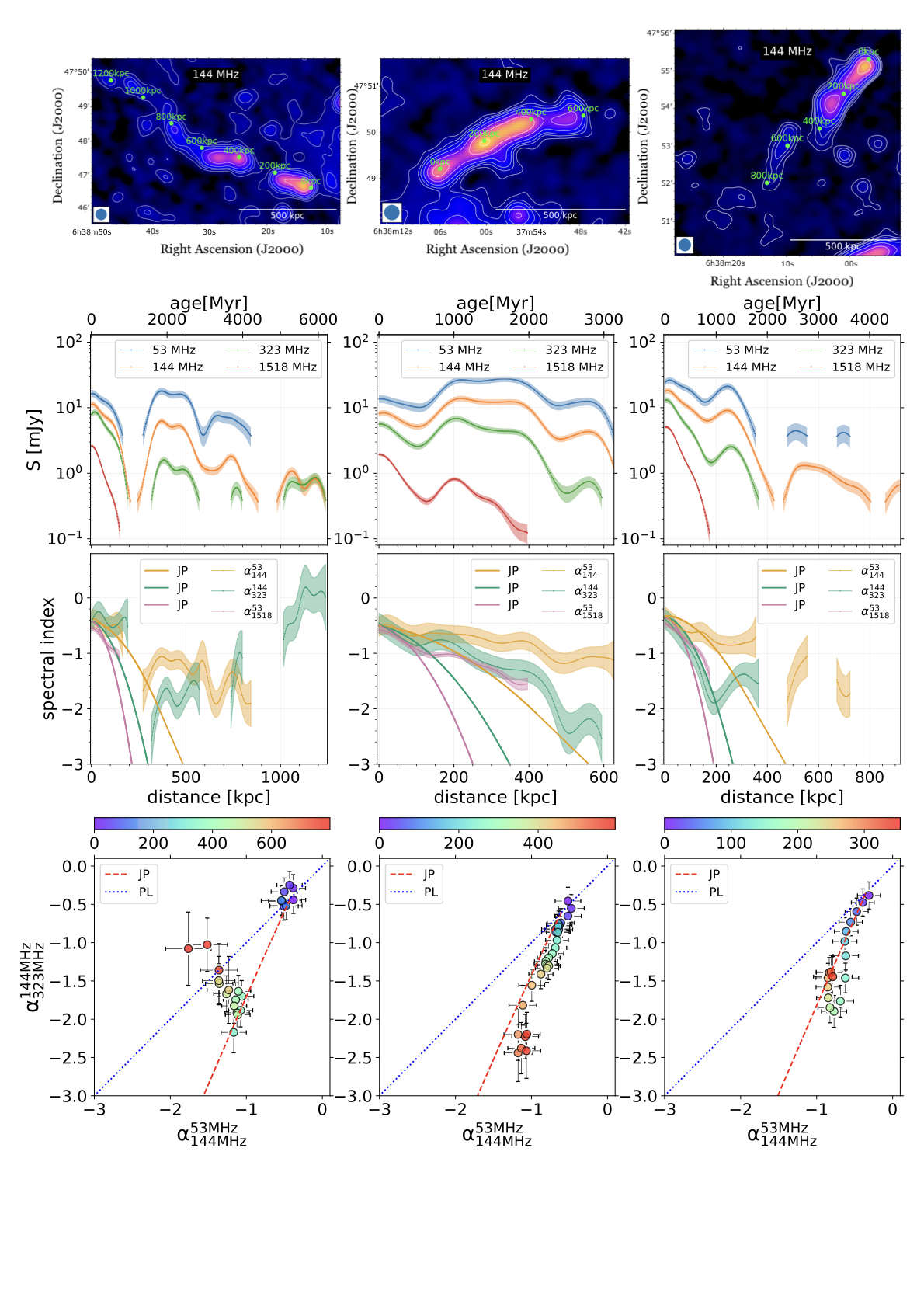}
     \vspace{-3.1cm}
     \caption{\textit{First row:} HT extension at 144 MHz, where the tails are more visible. \reviewfirst{Reference points are shown to help the reader in matching the profiles with images};  \textit{Second row:} flux profile; \textit{Third row:} spectral index profile considering regions with flux above $3\sigma_{\rm rms}$; \textit{Fourth row:} three frequency (53, 144, 323 MHz) color-color plot with the bisector indicating the set of point where $\alpha_{144}^{53} = \alpha_{323}^{144}$, which means that the spectrum resembles a power-law (PL). The dashed red line  represents the curved spectrum ($\alpha_{144}^{53} > \alpha_{323}^{144}$) of the JP model. The colorbar reflects the distance from the head of each point. 
    The profiles are shown for HT-A (first column), HT-B (second column) and HT-C (third column).
    }
\label{fig:tails_profiles}
\end{figure*}

In order to identify possible departures from a pure-ageing model (see \cref{subsec:JP} for details), we extracted flux densities and spectral index profiles (\cref{fig:tails_profiles}). The flux densities are extracted using beam-size sliding regions from the nucleus until the end of the tail, based on their largest linear extent at 144 MHz (\cref{fig:tails_profiles}, \textit{first row}). 
\reviewfirst{The sliding regions are separated by a few kpc from each other. In this way, we relaxed the requirement of independent measurements, which is not relevant in this case, as we do not use those values for fitting purposes.}
We then calculated the spectral index profiles between 54-144, 144-323 and 54-1518 MHz, considering only the regions with fluxes $\geq 3\sigma_{\rm rms}$ for each pair of frequencies (\cref{fig:tails_profiles}, \textit{second and third row}).
Another common method to analyse synchrotron spectra is through colour-colour plots \citep[e.g.][]{Katz-Stone1993, Stroe2013,Edler2022, Rudnick2022}. 
This method employs two, two-point power laws to characterize the curved shape of a source spectrum. For these plots, we use measurements made over beam-size separated regions of the source. The results are then compared to the expected trajectories in the color-color space, predicted by the standard model of spectral ageing (\cref{fig:tails_profiles}, \textit{fourth row}).
This method is useful to identify points that still follow the same spectral curvature even though they are spatially distant from each other.
In the next section, we describe our model and we study their behaviour separately.
%
%
%
\newpage

\begin{figure}
    \centering
    \includegraphics[width=\columnwidth]{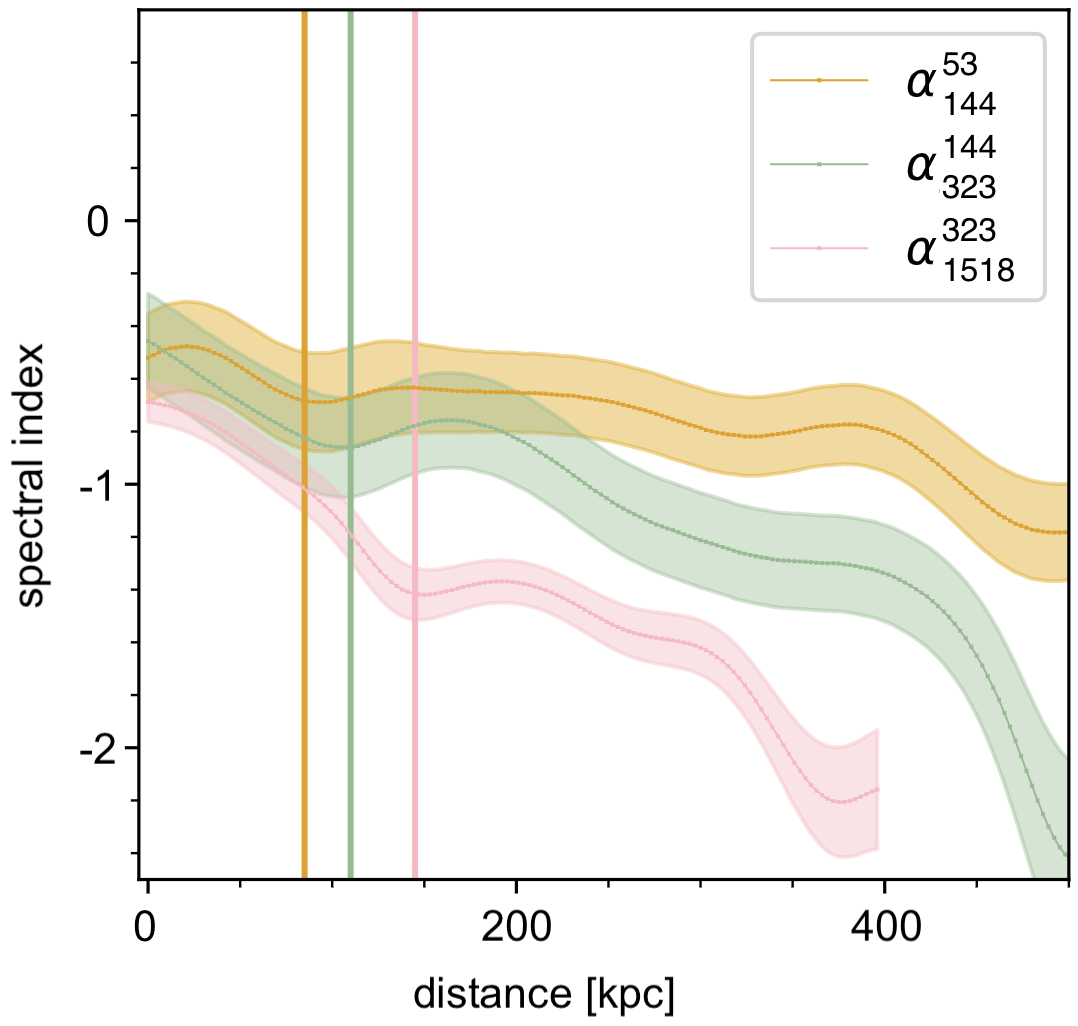}
    \caption{Spectral index profile for HT-B: comparison between $\alpha_{144}^{53}$, $\alpha_{323}^{144}$ and $\alpha_{1518}^{323}$. Vertical lines trace the shift of the flattening point from $\sim 150$ to $\sim 110$ and $\sim 80$ kpc.}
    \label{fig:tailB_spectral_ind_only}
\end{figure}

%
%

\subsection{Spectral ageing}\label{subsec:JP}
Once injected in the ICM from the AGN, the relativistic plasma starts loosing energy due to synchrotron and inverse Compton losses. 
At the point of the initial acceleration, the energy distribution of the electron population is assumed to be a power law in the form
$N(E) = N_0 E^{-\delta}$, where $\delta$ is the power-law index.
At later times $t>0$, we expect the energy distribution of the electron population to have a characteristic form depending on $t$. Since the characteristic time-scale for synchrotron losses is  $\propto 1/E$, the electron energy distribution at $t>0$ will be characterized by a depletion of the highest-energy electrons. The same reasoning applies to the synchrotron spectrum, which steepens with time away from the initial power law shape $S\propto\nu^{-\alpha}$ with $\alpha=(\delta-1)/2$. 
%
In the case of HT radio galaxies, this results in a spatial evolution of the spectrum from the head down to the tail. Specifically, the further from the nucleus, the older the electrons, i.e. they were accelerated at earlier times, and the more curved the spectrum is expected to be. In the absence of re-acceleration mechanisms, we expect to see the aged synchrotron spectrum assuming a characteristic curved shape.\\

Here we first consider a pure-ageing scenario to describe the flux and spectral index evolution. One of the most commonly used models to describe ageing of radio galaxies is the Jaffe-Perola \citep[JP;][]{JaffePerola1973} model. It accounts for ageing of plasma assuming:
\begin{enumerate*}[label=(\textit{\roman*})]
\item negligible adiabatic losses, in line with models where the tail flow is contained and channeled along magnetic field lines;
\item a single initial injection event with given injection index $\alpha_{\rm inj}$, that represent the plasma spectral index when it is first injected in the ICM by the AGN, i.e. before any ageing;
\item a given uniform magnetic field that we decide to keep fixed at the minimum aging value $B=B_{\rm min}$.  This choice means maximizing the lifetime of the emitting electrons by minimizing their losses due to synchrotron and IC: higher or lower values of the magnetic field would cause the radio source to fade faster. This equals to $B_{\rm min}=B_{\rm CMB}/\sqrt{3}$, where $B_{\rm CMB} = 3.2\times(1+z)^2 \,\mu$G is the equivalent magnetic field of the Cosmic Microwave Background at the source redshift, which for \Zwcl corresponds to $B_{\rm min} = 2.55\,\mu$G.
\end{enumerate*}
We calculated the spectra of each head in a region 1.5 times bigger than the beam, centered on the head (\cref{fig:heads_tail_spectrum}). We fix the injection index using the lowest frequencies available in this work \ainj $=$ \aLBAHBA, since they are the less affected by ageing (\cref{tab:tails_summary}). \\ 
In order to obtain the JP profiles we used the \textsc{pysinch}\footnote{\href{https://github.com/mhardcastle/pysynch}{https://github.com/mhardcastle/pysynch}} libraries \cite[presented in][ and subsequent work]{Hardcastle1998}. \reviewfirst{Through these libraries, we built an initial electron spectrum (with the injection energy index \dinj derived from the observed \aLBAHBA) and let it age in the minimum magnetic field condition, following the JP model. We then evolved the JP spectrum at $53,144,323,1518$ MHz and obtained the flux density (together with the spectral-index). We normalized the initial JP flux density to the first data point, which corresponds to the head of the tail.
}
The time-ageing of the JP model spectra can be converted to space-ageing along the tail, considering a given \reviewfirst{constant} velocity. We thus obtain a rough estimation of the projected velocity on the plane of the sky $v_{\perp}$ which is of the order of $\rm 200-300 \,km\,s^{-1}$ (\cref{tab:tails_summary}). We point out that this is a lower limit for multiple reasons. First, magnetic field and velocity are strongly correlated, which means that it is not possible to discern between the minimum ageing scenario and any other value of the magnetic field $B$, far from $B_{\rm min}$, combined with higher velocities of the tail.
Secondly, this velocity is projected, meaning that there could be a non-negligible component along the line of sight. 

\paragraph*{HT A} 
The radio galaxy HT A shows unusual properties: after the bright nucleus (the head), we note another bright peak in the surface brightness at $\sim$~380 kpc (\cref{fig:tails_profiles},  \textit{second rows}). Correspondingly, the spectral index profile shows a flattening, jumping from $\sim -0.5$ of the head to $\sim -1.5$ and then stabilizing at $\sim -1.3$ in both $\alpha_{53}^{144}$ and $\alpha_{144}^{320}$. The flattening extends to $\sim$~550 kpc (\cref{fig:tails_profiles},  \textit{third rows}).
We overplot the JP model (solid lines) for all frequencies. 
Up to 200 kpc, 
the spectral index profile decline as expected due to radiative losses of the electron population accelerated by the radio galaxy. In this first part, the JP model is able to reproduce the profiles. At larger distances, it is not trivial to account for the bump in the flux density profile and the flattening of the spectral index, using this pure-ageing scenario. 
Moreover, the color-color plot shows that points corresponding to 380-550 kpc are progressively moving from the expected curved JP towards the power-law line. That means the shape of the spectrum is getting flatter instead of continuing the steepening due to ageing.
Thus, we conclude a mechanism able to re-energise the relativistic population of CRe, initially injected by the AGN, must have occurred. 
We also note that there are few points in the range 740 - 800 kpc from the head that surprisingly cross the power-law line, lying in the area where $\alpha_{144}^{323} > \alpha_{53}^{144}$.
We argue this is an effect due to mixing of the electrons along the line of sight. Such an effect is common at very large distances from the AGN.

\paragraph*{HT B}

Using GMRT and \reviewfirst{VLA} observations, \cite{Cuciti2018} found a surface brightness jump at $\sim$ 150 kpc from the head of the tail, in correspondence with a flattening ($150-300$ kpc) of the spectral index. This was interpreted as evidence for electron re-energisation and they suggested that an interaction with a shock can explain the brightness and spectral index properties. The proposed scenario included the presence of a weak shock (Mach number $\mathcal{M} \lesssim 2 $), moving from the back end of the tail until the point of the brightness bump $\sim200$ kpc. The geometry of this scenario includes the tail travelling outward from the cluster centre with an angle 
$\theta_0=60^{\circ}$ to the line of sight.
This model was able to explain the high-frequency radio observations for reasonable choices of the model parameters, as for example the velocity of the tail of the order of $\rm \sigma_{v}\sim1400\,km\,s^{-1}$. Nonetheless, they did not detect any shock at the jump position in the X-ray image. Thus they concluded that, if the shock is present, it should be moving with a significant inclination with respect to the plane of the sky \citep[see Fig. 13 in][]{Cuciti2018}.

Our results at high frequencies are consistent with the brightening starting at $\sim 150$ kpc, and the subsequent spectral flattening.
As a reference with the analysis of \cite{Cuciti2018}, in \cref{fig:tailB_spectral_ind_only} we show $\alpha_{53}^{144}$ compared to  $\alpha_{323}^{1518}$.
The spectral profile shows a peculiar feature: the flattening starts at different distances depending of the frequencies used. To better appreciate this feature, we zoom in the first 400 kpc of HT-B and performed spectral index profiles comparing $\alpha_{144}^{53}$, $\alpha_{323}^{144}$ and $\alpha_{1518}^{323}$.
In \cref{fig:tailB_spectral_ind_only}, the flattening point appears to be spatially shifted from 150 kpc ($\alpha_{1518}^{323}$) to 110 kpc ($\alpha_{323}^{144}$) and 80 kpc ($\alpha_{144}^{53}$). 
This frequency-dependent feature would discourage the shock scenario, since the latter would create re-energisation feature happening simultaneously at the same point of the tail.
We discuss and test possible scenarios in \cref{sec:discussion}.
Interestingly, we also find a residual emission with $\alpha_{144}^{53} < -2.5$ at the very end of the HT B (\cref{fig:tails_allfreq}). This is clearly visible only in the LOFAR low-frequencies images and after a region where there is no emission. If this steep spot is part of the tail, the total extent of the HT galaxy would be of 1.2 Mpc. 
However, we cannot rule out that it may be a residual emission, also present in other parts of the image.
In addition, given the large distance and the significant spectral index uncertainty associated with it, we exclude the radio spot from further analysis.

\reviewfirst{
HT-B is the tail with the most extended frequency coverage. This means we can follow the flux density and spectral index evolution up to $\rm\sim 400~kpc$ at all the available frequencies in this work. }
\begin{figure*}

\centering
     \includegraphics[width=\textwidth]{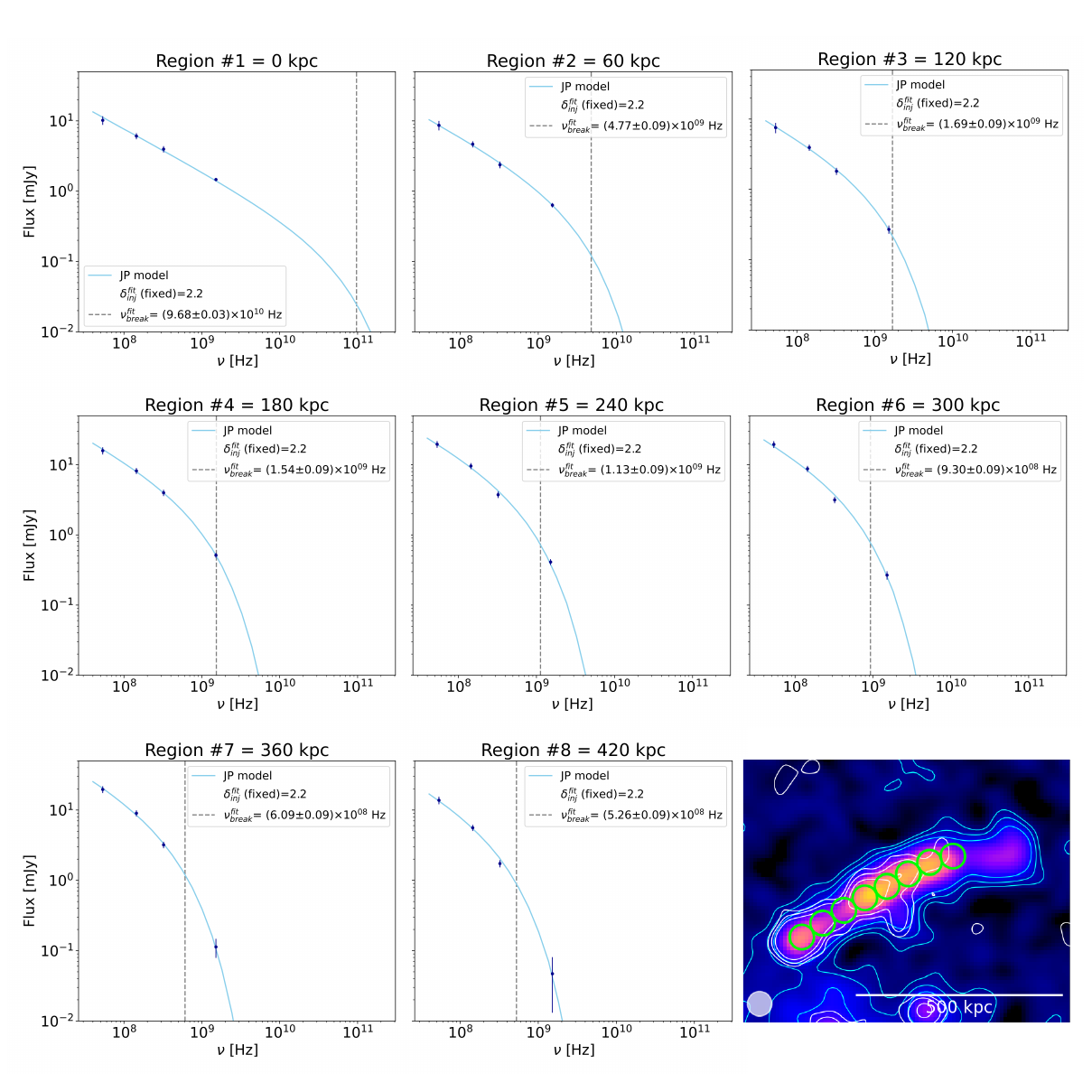} 
     \caption{
     \reviewsecond{Spectra extracted in 8 circular beam-sized independent regions along the HT-B. The light blue line represent the best fit of the JP model and the model parameters ($\delta_{\rm inj}$, $\nu_{\rm break}$) are listed in the legend of each plot. 
     The image of HT-B at 144 MHz is shown in the bottom right panel with the 8 regions overlapped. We plot the 3-, 6- and 12$\sigma_{\rm rms}$ contours at 144 (cyan) and 1518 (white) MHz, as a reference of the tail extension at different frequencies.}
     }
     \label{fig:local_spectra_JPfit}
\end{figure*}
\begin{figure*}
    \centering
    \includegraphics[width=\textwidth]{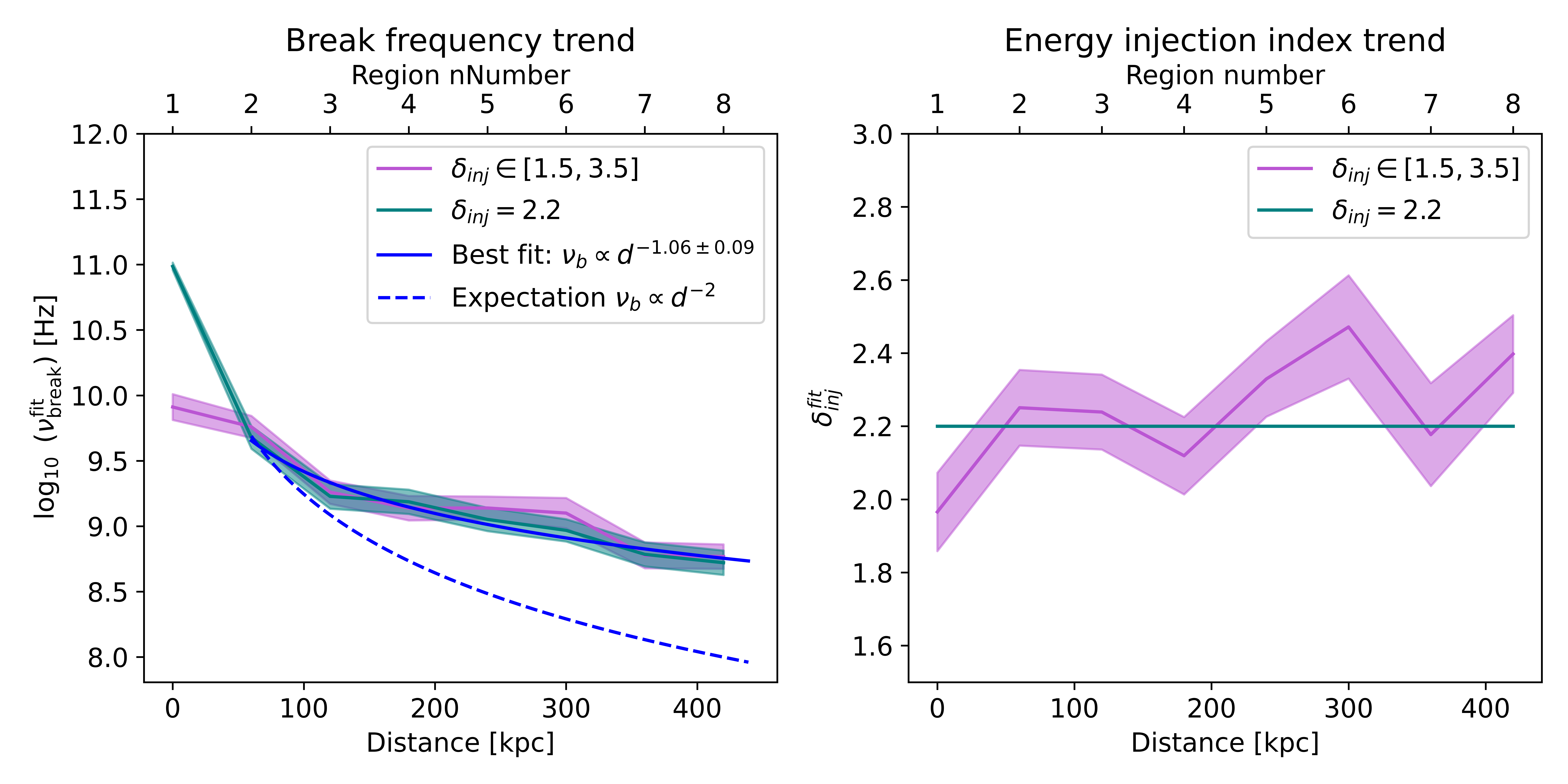}
    \caption{Best-fit break frequency and energy spectral index trend along HT-B. In purple the result with energy injection index free to vary,  $\delta_{\rm inj}\in [1.5,3.5]$; in teal it is kept fixed $\delta_{\rm inj}=2.2$ (i.e. $\alpha_{\rm inj} = 0.6$, in agreement with the HT-B head spectrum of \cref{fig:heads_tail_spectrum}).
    \reviewsecond{\textit{Left panel:}
    A power-law function was fitted to $\nu_{\rm break}^{\rm fit}$ corresponding to regions number 2 to 8. The best fit power-law index results to be $a=-1.06\pm 0.09$ (blue solid line). The break frequency trend expectation from a pure ageing scenario $\nu_{\rm break}\propto d^{-2}$ is plotted as a reference (blue dashed line).
    }}
    \label{fig:local_spectra}
\end{figure*}
\reviewfirst{
Therefore, in addition to the profiles shown in \cref{subsec:JP}, we performed local spectral fitting along HT-B. This allows us to study possible variations of the synchrotron spectrum along the tail. To do that, we extracted the spectra in eight circular beam-size independent regions along the source and fitted each local spectrum with a JP model using \textsc{synchrofit}\footnote{\href{https://github.com/synchrofit/synchrofit}{https://github.com/synchrofit/synchrofit}} package. \reviewsecond{The result of the local JP fitting is shown in \cref{fig:local_spectra_JPfit}}. The model has three free parameters: the energy injection index $\delta_{\rm inj}$, the break frequency $\nu_{\rm break}$ and the normalisation factor $N$ for correct scaling.
\reviewsecond{From the syncrhotron theory \citep[see e.g.,][]{Longair2011book}, the break frequency is related to the spectral age $t_s$ as
} 
\begin{equation}
\label{eq:nu_break}
    t_s = 1.61\times 10^3 \frac{B^{0.5}}{ (B^{2}+B_{\rm CMB}^{2}) (\nu_{\rm break}(1+z))^{0.5}} \quad \rm Myr
\end{equation}
where the magnetic field $B$ and the equivalent magnetic field of the CMB $B_{\rm CMB}$ are in $\mu G$, while $\nu_{\rm break}$ in GHz.
The initial energy spectrum spectrum undergoes a gradual transformation, with an exponential decline emerging beyond the point $\nu_{\rm break}$. 
If no source of re-acceleration is present, the measured break frequency decreases systematically along the tail \reviewsecond{ as $\nu_{\rm break}\propto t_s^{-2}$. In the approximation of constant velocity that leads to $\nu_{\rm break}\propto d^{-2}$}, in agreement with the scenario in which the oldest particles are those at a larger distance from the AGN.
We first fitted the spectra with the injection index free to vary in the range $[1.5,3.5]$ (purple, \cref{fig:local_spectra}). We noticed that the injection index does not show a clear trend but fluctuates around $2.2\pm0.2$. Thus we fitted again the same regions fixing the energy injection index to 2.2 (i.e. \ainj $=0.6$), which is also in agreement with the low-frequency power-law index of synchrotron emission, \aLBAHBA, observed of HT-B (\cref{fig:heads_tail_spectrum}). Thus we can limit the number of free parameters and we show the trend of the best-fit parameters $\nu_{\rm break}$ and $\delta_{\rm inj}$ in \cref{fig:local_spectra} (teal).
Keeping $\delta_{\rm inj}$ fixed or free to vary, does not influence the overall trend of the frequency break along the tail. Interestingly, we notice that after an initial decrease (up to region 3), the break frequency remains fairly constant. The distance at which this happens coincides with the distance of the spectral index flattening at $\sim 150$ kpc. This confirms that at some point along the tail, there must be a mechanism that stops the natural ageing (spectral index flattens and the break frequency stops decreasing).
}
\reviewsecond{To quantitatively assess the slowdown of the break frequency steepening, we employed a general power-law model of the form $f(d) \propto d^{-a}$ to fit the data points spanning from Region 2 to 8 (\cref{fig:local_spectra}, \textit{left}), where $d$ is the distance from the core and $a$ the power-law index. 
We excluded the break frequency value in Region 1, as is strongly vary among the two JP fitting models, and it can be considered as a lower limit since it is much higher than the highest spectral point in the case $\delta_{\rm inj}=2.2$.
The best fit (blue solid) shows a power-law index $a=1.06\pm 0.09$. Notably, this value indicates a slower rate of change compared to expected law based on the relationship between spectral age and break frequency (see \cref{eq:nu_break}) and assuming a constant tail speed, $\nu_{\rm break}\propto d^{-2}$ (dashed blue line for comparison). 
}
\reviewfirst{Similar evidence has been found in \citep{Parma1999}, who claimed re-acceleration processes for four of the radio galaxies in their sample.}

\paragraph*{HT C} 
Analysing HT C is difficult because of its morphology and extension. In fact, the radio galaxy is more compact at $\sim$ GHz frequency (see also \cref{fig:zoomin_tails}), while revealing a longer structure only in LOFAR images. 
Nevertheless, also in this case a pure-ageing JP model cannot account for the emission observed at distances larger than 200 kpc. 
Again, we see evidence for re-energised CRe left behind the AGN. These CRe do not age as expected by the considered model: the flattening of the spectral index implies that the energy distribution of the CRe is boosted toward higher energies.\\
%
%
%

\subsubsection{Other ageing models}
\label{subsec:KP}
Another widely used model for spectral ageing is the \cite{Kardashev1962} and \cite{Pacholczyk1970book} model (KP). 
The JP and KP models differ in the treatment of the emitting electrons' pitch angles, i.e. the angle between the magnetic field $\textbf{B}$ and velocity $\textbf{v}$ of the charged particle. 
In the KP model, the pitch angles of the electrons are fixed over their radiative lifetime, while in the JP model, pitch angle scattering takes place. This difference influences the way CRe lose energy, and thus, affects the resulting synchrotron spectra \citep{Harwood2013}.
The pitch angle scattering of the JP model is a more likely scenario than that of the fixed pitch angles of the KP model \citep{Tribble1993}. The reason for this is that for a turbulent magnetic field configuration, the expected resonant scattering of the particles is effective on scales comparable to the Larmor radius of CRe, and thus smaller than the resolution of observations \citep{Carilli1991, Hardcastle2013}.
Finding the best model for radiative ageing is beyond the scope of this work (see \cref{sec:discussion}). We note that the differences between JP and KP model spectra would not change our conclusion, i.e. standard ageing cannot account for the observed profile and thus other re-acceleration mechanisms need to operate in this system.

\subsection{X-ray properties}
\label{subsec:X-ray_properties} 

We use XMM-Newton observations to study the environment of the HT galaxies, i.e. the ICM that deflects their jets. From previous studies, it is known that this cluster appears to be in a non-relaxed state \citep{Cuciti2015}. 

\subsubsection{Surface brightness profile}
\label{subsubsec:SB_X-ray_emission}

\begin{figure}
    \centering
    \includegraphics[width=\columnwidth]{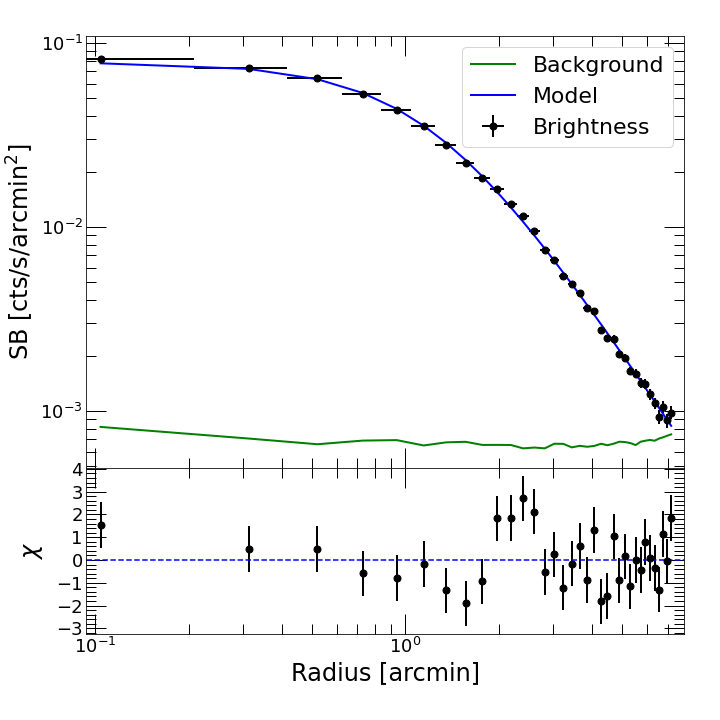}
    \caption{Surface brightness profile with the fitted single $\beta$-model. The best fit parameters are $\beta=0.59\pm 0.05$, $r_c=1.24\pm 0.02$ arcmin, $\rm I_0=0.079\pm 0.006\, cts/s/arcmin^2$, with a reduced $\chi ^{2} = 1.2$.
    }
    \label{fig:single_beta}
\end{figure}

We first studied the X-ray surface brightness to check for jumps in the azimuthal profile. We used \texttt{pyproffit\footnote{\href{https://github.com/domeckert/pyproffit}{https://github.com/domeckert/pyproffit}}} \citep{Eckert2020Opyprofit} to compute and model the surface brightness with a single, spherically symmetric $\beta$-model  \citep{CavaliereFusco1976} of the form:

\begin{equation}
    I(r) = I_0  \Big[ 1+ \Big( \frac{r}{r_{\rm c}}\Big)^2 \Big]^{-3\beta+0.5} + b ,
    \label{eq:beta_model}
\end{equation}
where $I_0$ is the central surface brightness, $r_{\rm c}$ the core radius, $\beta$ describes the ratio between thermal and gravitational energy of the plasma and one additional parameter $b$ to describe the background \citep{Eckert2020Opyprofit}.
We extracted a profile in circular annuli centered on the X-ray peak, with a bin size of $15\arcsec$ until $R_{500}$. The azimuthal fit shows that a smooth $\beta$-model accounts for the emission, which does not show bumps or discontinuities (\cref{fig:single_beta}).

\subsubsection{Gaussian Gradient Magnitude Filter}
\label{subsec:X-ray_GGM}

\begin{figure}
    \centering
    \includegraphics[width=\columnwidth]{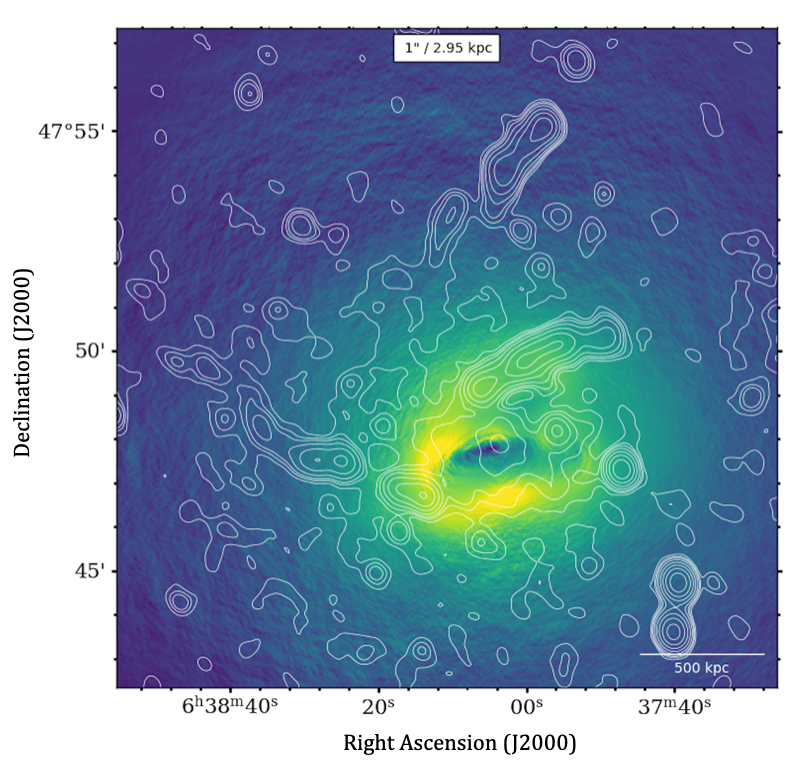}
    \caption{X-ray 0.5 - 2 keV image filtered with an adaptive GGM filter, using a signal-to-noise ratio of 32. Radio contours from \cref{fig:tails_HBA} are overplotted.}
    \label{fig:GGM_adaptive}
\end{figure}

In order to look for  features in the ICM that can be connected to the HT galaxies, we used standard X-ray image processing techniques. 
When studying galaxy clusters, particularly relaxed objects, the bright core of ICM emission gives rise to steeply peaked surface brightness profiles, where it can be difficult to detect variations such as edges, filaments, cavities or ripples.
Thus, filtering techniques are frequently used to see and enhance variations in  the surface brightness distribution. Among the different techniques, we used the adaptive version of the Gaussian Gradient Magnitude\footnote{\href{https://github.com/jeremysanders/ggm}{https://github.com/jeremysanders/ggm}}  \citep[GGM;][]{Sanders2016MNRAS457,Sanders2016MNRAS460} filter.
In standard GGM filter, the image is convolved with a function that is the gradient of a Gaussian with a particular scale, measuring the gradients on this scale. This is a useful method to detect edge-like structures, such as shocks and cold fronts, which lead to strong gradients \citep{Sanders2018ChandraNews}. Since the accuracy of the gradient depends on the number of counts within the scale probed, this standard GGM filtering becomes noisier in regions where the count rate is lower, i.e. the outskirts of galaxy clusters. Therefore, we used the adaptively smoothed GGM described in \cite{Sanders2022} that dynamically choses the appropriate scale that reduces this effect. The main input parameter of this software is the signal-to-noise ratio, which is used to calculate the smoothing scale from the input counts.
Following \cite{Sanders2022} we set it to 32, and we used the image counts in the energy band from 0.5 to 2 keV. 
We note the presence of bright regions, i.e., high gradient values, in the central left part of the GGM-filtered image (\cref{fig:GGM_adaptive}) but this region does not intersect any of the tails. Hence, we conclude that the ICM does not show features that appear to interact with the radio tails.

\subsection{Radio halo morphology}\label{subsec:radio_halo} 

Based on the 3$\sigma_{\rm rms}$-contours, the radio halo in \Zwcl has a diameter of $\sim$ 400-800 kpc in the E-W direction, depending on the frequency (see \cref{tab:H_summary}). 

Following the procedure of \cite{Cuciti2018}, we measure the flux density within a circular area of $\sim$ 600 kpc diameter, which roughly follows the $3\sigma_{\rm rms}$-contours of the images (see \cref{fig:halos_allfreq}), using the source-subtracted tapered images described in \cref{subsec:sources_subtraction}. This region avoids the two HT-A and HT-B radio galaxies, which are extended sources, and thus cannot be subtracted from the \textit{uv}-data with the same procedure used for the compact sources in \cref{subsec:sources_subtraction}. 
The results are listed in \cref{tab:H_summary}. We point out that we find high-frequency flux densities that are slightly different from \cite{Cuciti2018}. Our values of $S_{\rm 1.5GHz}=2.5\pm0.2$ mJy and $S_{\rm 320~MHz}=14\pm2$ mJy are nevertheless still consistent with measurements reported in \cite{Cuciti2018}. The difference could be due to different sources subtraction methods. \cite{Cuciti2018} measured the compact-sources flux density from the high-resolution images and subtracted that from the measurement on the radio halo region of the low resolution images, while here we subtracted the compact sources directly from the $uv$-data.

In order to obtain a size and flux estimate that is independent of the signal-to-noise ratio of the radio images (i.e., that does not rely on the $3\sigma_{\rm rms}-$contours), we fit the the same low-resolution source-subtracted images of \cref{fig:halos_allfreq} with Flux Density CAlculator software\footnote{\href{https://github.com/JortBox/Halo-FDCA}{https://github.com/JortBox/Halo-FDCA}} \citep[\texttt{FDCA-Halo};][]{Boxelaar2021}. This method fits the surface brightness profile of halos to a 2D-model using Bayesian inference.
The general exponential model used is
\begin{equation}
\label{eq:FDCA_exponential_profile}
    I(\textbf{r}) = I_0e^{-G(\textbf{r})} ,
\end{equation}
where $I(\textbf{r})$ is the surface brightness and $G(\textbf{r})$ the function that takes different forms depending on the complexity of the model \citep[see][for details]{Boxelaar2021}. In the circular model case, $G(\textbf{r})=|{\textbf{r}|/\re}$, where the e-folding radius $\re$ is the length-scale at which the surface brightness drops to $I_0/e$. 
To avoid contaminating sources, we first masked the regions corresponding to extended radio sources other than the radio halo. Masked regions, model and residuals can be found in \cref{appendix:FDCA}, while the fitting results in \cref{tab:H_summary}. 

The fitted $\re$ values are more uniform across the frequency range $\rm (120-145\,kpc)$ compared to the size based on the $3\sigma_{\rm rms}$-contours $(D \sim 480-850\,\rm kpc)$ and consistent with each other because they are less affected by the noise.
Flux densities with \texttt{FDCA-Halo} are integrated up to a radius $R = 3\re$ that contains 80\% of the total flux $S_{\nu}(<3\re)=0.8S_{\nu}^{\infty}$ (\cite{Murgia2009}). 
 
To date, this is one of the few radio halos where we can study the radio spectrum from 1.5 GHz down to 50 MHz. 
In the near future, observations by the LOFAR LBA Sky Survey \citep[LoLSS;][]{deGasperin2021,deGasperin2023}, combined with archival data from other telescopes (such as uGMRT and VLA), will increase the number of systems with such a large frequency coverage (Pasini et al., in prep.).

\begin{figure}
	\includegraphics[width=\columnwidth]{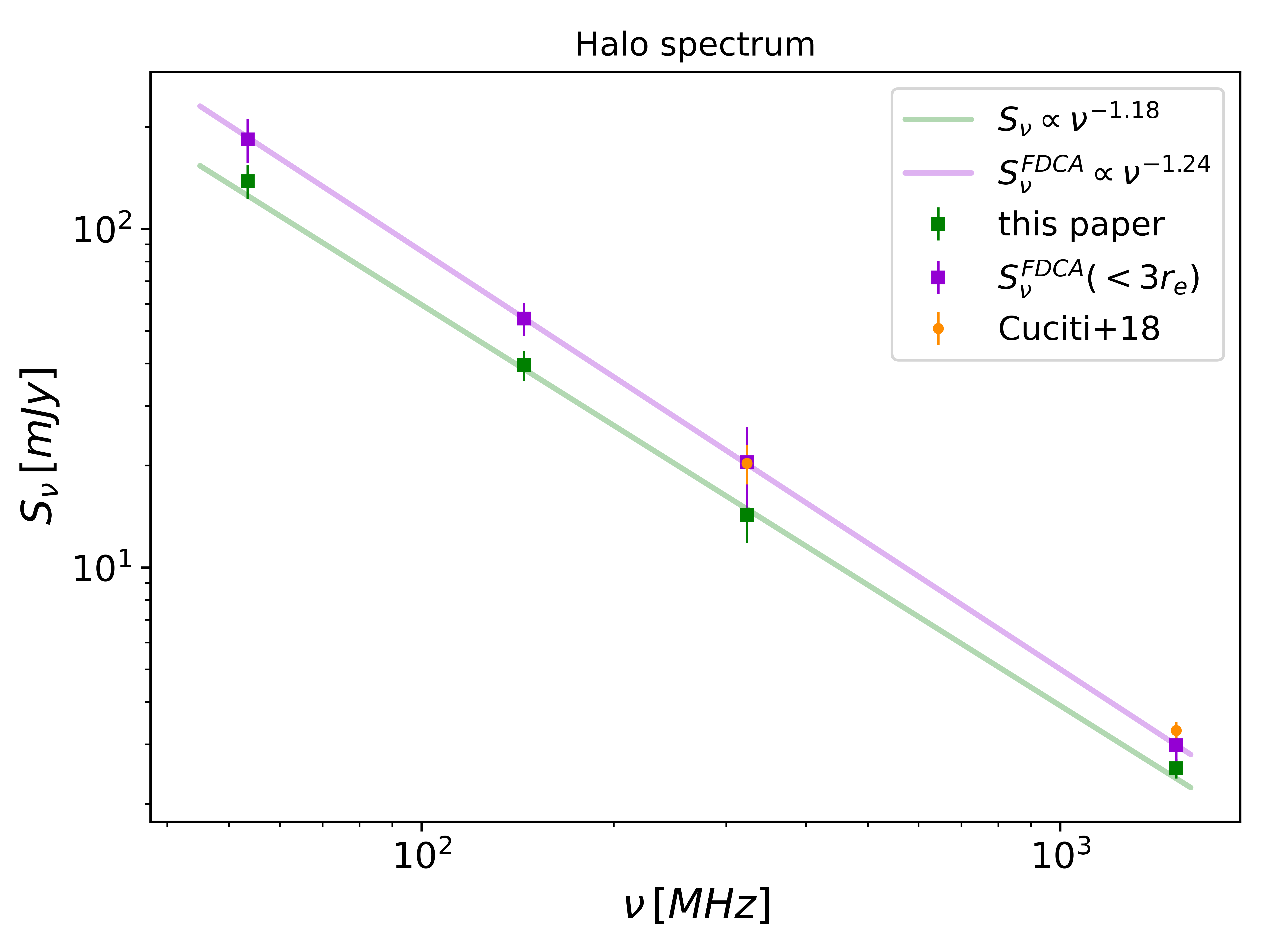}
    \caption{Integrated spectrum of the radio halo from 1.5 GHz down to 53 MHz. 
    Green points indicate the integrated flux values measured in a circular region with $R=300$ kpc. 
    Purple points represent $S_{\nu}^{\rm \scriptscriptstyle FDCA}(<3 \rm{r_{e}})$, the flux calculated via \texttt{FDCA-Halo} fitting procedure within three $e$-folding radii. 
    Values at 320~MHz and 1.5~GHz from \citet{Cuciti2018} are displayed in blue.}
    \label{fig:radiohalo_spectrum}
\end{figure}

The spectrum of the radio halo in \Zwcl is shown in \cref{fig:radiohalo_spectrum}: green points represent flux densities  extracted within a radius of $300$ kpc $S_{\nu}(<300\,\rm kpc)$, while purple points are values obtained via \texttt{FDCA-Halo} procedure $S_{\nu}^{\rm \scriptscriptstyle FDCA}(<3\re)$. We note that the fitted size of the halo, i.e. $3\re$, is wider than the circular extraction region with $R=300$ kpc (based on the $3\sigma_{\rm rms}-$contours), and thus the \texttt{FDCA-Halo} flux is higher. Despite of this, the spectral shape remains unchanged (\cref{fig:radiohalo_spectrum}). We fit the spectrum using a power law, finding a best-fit spectral index between 53~Hz and 1518~Hz of $\alpha=-1.18\pm0.04 $ and $\alpha^{\rm \scriptscriptstyle FDCA}=-1.24\pm0.03$, respectively. These values are consistent with previous high-frequency \citep{Cuciti2018} and low-frequency \citep{Cuciti2022Nat} studies.
Spectral indices $\alpha < - 1.1$ are expected for the radio halos and are in agreement with other radio halos studied over the whole radio spectrum (e.g., A1758 \cite{Botteon2018A1758, Botteon2020A1758}, Coma \cite{Bonafede2022}).
Moreover, this halo is under-luminous in the power-mass correlation in both $P_{\rm 1.5GHz}-M_{500}$ \citep{Cuciti2021} and $P_{\rm 150MHz}-M_{500}$ (Cuciti et al., submitted); and has an emissivity $J_{\rm 1.4 GHz}=(3.2\pm0.2)\times 10^{-43}$ erg/(s \,cm$^3$Hz) slightly lower compared to the emissivity of classical radio halos at 1.4 GHz in the range of [5$\times 10^{-43}$; 4$\times 10^{-42}$]\,erg/(s \,cm$^3$Hz) \citep{Cuciti2022Nat}.
Statistical studies showed that halos not lying on the $P_{\rm 1.5GHz}-M_{500}$ correlation are usually associated with non-merging clusters or with USSRHs. In the framework of the turbulent re-acceleration scenario, this means that either the radio halo is not present at all, or the system is less massive and has experienced minor mergers, thus producing radio halos with steeper spectra.

\begin{table*}
\small
\begin{center}
\renewcommand\arraystretch{1.5}
\caption{\label{tab:H_summary} Summary of the main radio halo properties}
\begin{tabular}{ c c c c c c c c}
\hline
\hline
    $\nu$   & $ D $          & $S_{\nu}(<\rm 300\,kpc) $     &   $P_{\nu}$      & $I_0$ & $\re$ & $S_{\nu}^{\rm \scriptscriptstyle FDCA}(<3\rm{r_{e}})$ & $P_{\nu}^{\rm \scriptscriptstyle FDCA}$\\ 
 $\mathrm{MHz}$& $\mathrm{kpc}$ & $\mathrm{mJy}$    &   $\mathrm{W/Hz}$   &  $\rm \mu \mathrm{Jy}/ arcsec^{2}$  & $\mathrm{kpc}$ & $\mathrm{mJy}$ & $\mathrm{W/Hz}$\\ 
\hline
  $53$   & 770 &  $138\pm16$   & $(1.2\pm1)\times 10^{25} $    & $16\pm2$      & $141\pm14$ & $184\pm27$   &  $(1.6\pm0.2)\times10^{25}$   \\ 
  $144$  & 850 &  $40\pm4$     & $(3.5\pm0.3)\times 10^{24} $  & $4.5\pm0.3$   & $145\pm7$  & $54\pm6$     &  $(4.7\pm0.2)\times10^{24}$   \\ 
  $320$  & 480 &  $14\pm2$     & $(1.2\pm0.2)\times 10^{24} $  & $2.1\pm0.5$   & $130\pm28$ & $20\pm6$     &  $ (1.8\pm0.5)\times10^{24}$   \\ 
  $1518$ & 530 &  $2.5\pm0.2$  & $(2.2\pm0.2)\times 10^{23} $  & $0.36\pm0.04$ & $120\pm12$   & $3.0\pm0.3$  &  $ (2.6\pm0.3)\times10^{23} $  \\ 
 \hline
\end{tabular}
\end{center}
\begin{enumerate*}[start=1,label={Col. \arabic*:},align=left]
    \item Frequency;
    \item Linear extension corresponding to the $\rm3\sigma_{\rm rms}$ contour in the E-W direction as a size reference;
    \item Flux density measured within a circular region with $R=300$~kpc (see \cref{fig:halos_allfreq});
    \item Radio power calculated from $S_{\nu}(<300$ kpc) with the K-correction applied, assuming $\rm \alpha = -1.2$;
    \item Central surface brightness from \texttt{FDCA-Halo} fitting, using a circular halo model;
    \item e-folding radius from \texttt{FDCA-Halo} fitting, using a circular halo model;
    \item Flux density measured within the $3r_{e}$ calculated by \texttt{FDCA-Halo};
    \item Radio power calculated by \texttt{FDCA-Halo} corresponding to  $P_{\nu}^{\rm \scriptscriptstyle FDCA}(<3\rm{r_{e}})$, assuming $\rm \alpha = -1.2$
    \textcolor{white}{----------------------------------------------------}.
\end{enumerate*}
\end{table*}

\section{Discussion}
\label{sec:discussion}
\reviewfirst{
In this work, we studied HT radio galaxies that show ripples in the surface brightness profiles and spectral index flattening along the tail.
In order to assess whether this is in line with pure radiative ageing due to IC and syncrothron losses, we compared the data to the JP model. This is a standard and widely used spectral ageing model, that assumes electrons are injected with a constant power-law energy spectrum.
Moreover, in this analysis, we work under the assumptions of uniform minimum magnetic field, constant velocity and fixed inclination of the radio galaxy (see \cref{subsec:JP}). Though simple, these are reasonable hypotheses given, for example, that the tails do not show sudden changes in direction. More complex models that include variations of the magnetic field etc. would require more data.}
Our results show that the jet emission deviates from the pure radiative ageing due to IC and syncrothron losses. We conclude that this is evidence for re-acceleration.
In this section we discuss possible scenarios that can affect the tail's spectral behaviour and morphology. Particular focus is on the long flattening of the spectral index and a frequency-dependent feature, only clearly visible in HT B (\cref{fig:tailB_spectral_ind_only}). This is due to the lack of high frequency information to use for measuring the spectral index  $\alpha_{323}^{1518}$ at distances $>150-200$ kpc from the head for HT A and HT C.

A first simple explanation for ripples in the brightness profile can be a varying output of the AGN. The standard JP model (\cref{subsec:JP}) uses a single injection event. Thus, we are assuming that the AGN injection rate remains constant with time. %
A varying AGN would reproduce the ripples in the surface brightness profiles but not the flattening in the spectral index that we see in \cref{fig:tails_profiles}. 
Thus, although a varying AGN is still a plausible mechanism that contributes to the observed profiles, we exclude it as the main mechanism creating the observed features.
\reviewfirst{We point out that possible changes of inclination during the motion through the cluster can also alter the observed surface brightness and spectral index profiles, causing deviations from the expected JP profile. However, we argue that this scenario is very unlikely for two main reasons. 
First, the morphology of the tails does not show sharp bends that could suggest such a scenario of varying direction of motion. 
Secondly, even if projection effects happened along the tail, in those regions we would be summing younger “foreground” electrons with older “background” ones. The result would be a spectral index decreasing faster with respect to the standard ageing, not slower, as we observe.
}

\subsection{Shock - tail interaction}
\label{subsec:shock_tail}

The interaction between shocks and tails is a possible scenario that can alter the tail morphology and physical properties.
It is well known that when a shock encounters a population of aged electrons, it can re-energize particles, leading to observable synchrotron emission. However, the expected very high sound speed of the relativistic plasma does not allow the shock to penetrate into the radio tail \citep{EnsslinKrishna2001}. 

However, a shock can adiabatically compress the fossil plasma and increase the magnetic field \citep{EnsslinKrishna2001, EnsslinBrueggen2002}. In fact, adiabatic compression is able to re-energize an electron population, inducing brightening and spectral flattening. 
This kind of interaction was already investigated in the past, either in combination with the presence of a shock \citep{vanWeeren2017NatAs, Botteon2019, Wilber2019}, or a cold front \citep{Botteon2021,Giacintucci2022}.
The shock scenario was also proposed by \cite{Cuciti2018} as a possible explanation to the HT-B observed radio properties.
There are very few examples of HT-radio galaxies formation \citep{O'Neill2019_884}  or shock-radio galaxy interaction \citep{Nolting2019_876Aligned, Nolting2019_885Orthogonal, ONeill2019_887} simulations in the literature that can shed light on this complicated scenario. Although none of these simulations are tuned to reproduce the geometrical configuration proposed by \cite{Cuciti2018}, we can use them for a general comparison. 
In particular, \cite{Nolting2019_876Aligned} studied the interaction between a shock and an active radio galaxy, in the case where jets are aligned with the shock normal.  They show that, as the shock travels along the jets, vortices are generated behind it (see Fig. 5 of \cite{Nolting2019_876Aligned}). This toroidal vortex was found also in other simulations of a shock wave interacting and adiabatically compressing a radio plasma cocoon \citep{EnsslinBrueggen2002, PfrommerJones2011}.
The fact that we do not observe this morphological feature suggests that there is no shock travelling through the tail. However, we cannot completely rule out this option since signs of jets disruption might be present on scales smaller than our resolution.  Moreover, we point out the need of tailored simulations, where a shock crosses a head-tail radio galaxy whose morphology has been already modified by the interaction with the ICM.
As mentioned in \cref{subsec:JP}, a shock interacting with the tail in the same configuration proposed by \cite{Cuciti2018} would
create re-acceleration features at the same point of the
tail (i.e., simultaneous brightness jumps and flattening of the spectra in correspondence of the shock location), while in HT-B we observe a shift of the point where the spectrum flattens (\cref{fig:tailB_spectral_ind_only}). This will be discussed in detail in \cref{subsec:turbulence_tail}. Finally, we point out that the XMM-Newton observation does not show any shocks at or near the radio tails, (see \cref{subsec:X-ray_properties}). For all the arguments mentioned above, we consider the presence of a shock as an unlikely scenario to explain the data.

\subsection{Turbulence}
\label{subsec:turbulence_tail}

Turbulence within the tail can arise from hydrodynamic instabilities that result from the interaction between the relativistic plasma and the surrounding medium.  \citep{Mignone2004, Jones2017, Mukherjee2021, Kundu2021, Ohmura2023}. This turbulence can trigger several mechanisms of stochastic acceleration of particles, such as gentle acceleration, which has been suggested as a possible explanation for the GReET in Abell 1033 \citep{deGasperin2017MNRAS, Edler2022}.

As mentioned in \cref{subsec:JP}, data for HT-B suggest that the spectral index flattens at different distances, depending on the frequency. In particular, the spectrum measured at higher frequencies flattens at greater distances from the head. Considering the kinematics of the tails, this corresponds to a situation in which the spectrum at higher frequencies flattens over longer times. 
This scenario agrees with the fact that spectral flattening is more evident at higher frequencies: $\alpha_{53}^{144}$ remains almost constant in the first $400$ kpc around a value of $-0.6,-0.8$, while $\alpha_{323}^{1518}$ shows a well-defined decrease up to $\sim-1.4$ before flattening (see \cref{fig:tailB_spectral_ind_only}).
This evidence allows us to understand something more about the mechanisms of particle re-acceleration. 
In practical terms, the flattening of the spectrum at higher frequencies over longer times indicates a gradual shift of the particles towards high energies. 
This, in turn, places constraints on the re-acceleration times.
\textcolor{black}{
\cref{fig:tailB_spectral_ind_only} suggests that the energy of the emitting electrons essentially doubles in about 80 kpc, corresponding to about 260 Myr (considering a speed of the tail $\sim\rm 300 \,km/s$).
}
We can conceive of two scenarios. In the first scenario, the electrons cool due to losses for a time $t_{\rm start}$ without being re-accelerated. After a time $t_{\rm start}$, necessary for the development of the instabilities in the head tail and possibly also necessary for the turbulence to cascade to resonant scales, the same particles are re-accelerated on a time scale, $\tau_{\rm acc}$.
This causes the spectrum to increase again at higher energies if $\tau_{\rm acc}$ < $t_{\rm start}$ ($\tau_{\rm acc}\geq t_{\rm start}$ would otherwise simply balance cooling on time scales $\tau_{\rm acc}$). 
In this case, $\tau_{\rm acc}\sim 260 $ Myr and $t_{\rm start} \sim 500$ Myr (from kinematics $\sim$ distance spectral flattening/velocity).
A second option could be that a second component of (relativistic) particles is re-accelerated starting at $t_{\rm start}$ with $\tau_{\rm acc} \sim 200-300$ Myr. These particles might be present in the tail at lower energies and potentially filling a different volume with respect to the other component. This second scenario could explain the strong brightness increase observed in the re-acceleration phase and might also suggest that volume is not uniformly filled with CR particles.
Discriminating between the two possibilities and constraining model parameters by fitting the data is beyond the scope of the current work and will deserve a future study.
Finally, we point out that galaxies in clusters are expected to move with a velocity of the order of the velocity dispersion ($\sim 10^3$ km/s), which is not compatible with the value of a few hundreds km/s, derived from our modelling in \cref{subsec:JP}. 
This suggests deficits in the JP model, even in the first kpc from the AGN. This means that no part of the tail is ageing subject to pure radiative synchrotron and IC losses, and we are facing a mechanism that acts broadly along the tail and starts as soon as the jets interact with the external medium.
Recent simulations by \cite{Ohmura2023} show the effect of turbulent reacceleration on HT galaxies. They simulated AGN jets in an ICM wind and computed the evolution of CRe, considering both energy losses and stochastic acceleration. Their work clearly shows that in the presence of re-acceleration, the spectral index does not decrease substantially along the tail but remains fairly flat ($\alpha^{150}_{600}\sim-0.8, -1$) for about 150 Myrs. This is in agreement with our observations at low frequencies. At higher frequencies, we find a decrease of the spectral index in the first kpc from the head, which we attribute to the factors described above.

\section{Conclusions}\label{sec:conclusion}
In this work, we have studied the radio halo and three HT radio galaxies of the same cluster \Zwcl (z=0.174). We make use of LOFAR LBA 53 MHz, LOFAR HBA 144 MHz, GMRT 323 MHz, \reviewfirst{VLA} 1.5 GHz and X-ray XMM-Newton data. 
The cluster has already been studied at higher frequencies \citep{Cuciti2018}. However, the addition of low-frequency LOFAR observations has enabled us to study low brightness emission on large scales, both, in terms of diffuse halo emission and radio galaxies.\\
Our main results can be summarized as follows:
\begin{itemize}
    \item We studied three HT radio galaxies, spread over the whole cluster volume. Their properties are listed in \cref{tab:tails_summary}.  This is the first time where multiple tails belonging to the same cluster can be studied over a frequency range from 53 to 1518 MHz. The use of very low frequency observations allowed us to observe them up to $\sim$ 1 Mpc projected size and study the spectra at very large distances from the head.
    \item We investigated the properties of the HT galaxies by extracting surface brightness and spectral index profiles, revealing an increase of surface brightness and flattening of the spectral index.
    These characteristics together with their considerable lengths point towards a process that involves the re-acceleration of the initial electron population, which otherwise would cool below energies capable of radiating at our reference radio frequencies.
    \item We have proposed a scenario that could explain the morphological and spectral properties of the HT galaxies. This scenario involves a turbulence-induced, gentle re-acceleration mechanism that could explain the long flattening of the spectral profiles and the frequency-dependent feature of \cref{fig:tailB_spectral_ind_only}. 
    We propose two scenarios: one involving a standard single population of electrons which initially cools down and then starts to be re-accelerated after $t_{\rm start}$; a second one, including multiple components of relativistic particles that are activated at different times.
    Regardless of the scenario, the re-acceleration mechanism would act on a characteristic time scale $\tau_{\rm acc}\sim 200-300$ Myr or shorter, derived by comparing the distances at which the spectrum at different frequencies flattens. 
    \item \Zwcl also hosts a radio halo with a linear size varying from $\sim$400 to 800 kpc, depending on the frequency.  We find that the spectrum is well represented by a power law with a spectral index between 53 and 1518 MHz of  $\alpha^{\rm \scriptscriptstyle FDCA}=-1.24\pm0.03$ (\cref{fig:radiohalo_spectrum}). In previous studies, this halo was found to be under-luminous in both the power-mass correlation at 1.5 GHz and 150 MHz. Together with the relatively flat spectrum and the small size, we suggest that minor mergers can cause the extended radio emission. 
\end{itemize}
Our results show that low-frequency observations are key to observe phenomena producing low brightness and steep spectrum emission. The presence of multiple HT radio galaxies with extreme lengths suggest that radio galaxies in clusters can enrich the ICM with fossil electrons. 
Moreover, if the proposed gentle turbulent scenario is a common process for radio galaxies in clusters, then this would imply that large amounts of electrons released into the ICM by AGN would be able to survive with high energies producing a seed population of energetic particles. These seed electrons have been invoked to explain cluster-scale radio emission, such as radio halos and relics.

\section*{Acknowledgements}

MB acknowledges support from the Deutsche Forschungsgemeinschaft under Germany's Excellence Strategy - EXC 2121 "Quantum Universe" - 390833306.
CS acknowledge support from the German Federal Ministry of Economics and Technology (BMWi) provided through the German Space Agency (DLR) under project 50 OR 2112.
RJvW acknowledges support from the ERC Starting Grant ClusterWeb 804208.
GDG acknowledges support from the Alexander von Humboldt Foundation.
AB acknowledges financial support from the European Union - Next Generation EU.
LOFAR \citep{vanHaarlem2013} is the Low Frequency Array designed and constructed by ASTRON. It has observing, data processing, and data storage facilities in several countries, which are owned by various parties (each with their own funding sources), and that are collectively operated by the ILT foundation under a joint scientific policy. The ILT resources have benefited from the following recent major funding sources: CNRS-INSU, Observatoire de Paris and Université d'Orléans, France; BMBF, MIWF-NRW, MPG, Germany; Science Foundation Ireland (SFI), Department of Business, Enterprise and Innovation (DBEI), Ireland; NWO, The Netherlands; The Science and Technology Facilities Council, UK; Ministry of Science and Higher Education, Poland; The Istituto Nazionale di Astrofisica (INAF), Italy.
This research made use of the Dutch national e-infrastructure with support of the SURF Cooperative (e-infra 180169) and the LOFAR e-infra group. The Jülich LOFAR Long Term Archive and the German LOFAR network are both coordinated and operated by the Jülich Supercomputing Centre (JSC), and computing resources on the supercomputer JUWELS at JSC were provided by the Gauss Centre for Supercomputing e.V. (grant CHTB00) through the John von Neumann Institute for Computing (NIC).
This research made use of the University of Hertfordshire high-performance computing facility and the LOFAR-UK computing facility located at the University of Hertfordshire and supported by STFC [ST/P000096/1], and of the Italian LOFAR IT computing infrastructure supported and operated by INAF, and by the Physics Department of Turin university (under an agreement with Consorzio Interuniversitario per la Fisica Spaziale) at the C3S Supercomputing Centre, Italy.

\section*{Data Availability}

Raw data are publicly available in the archives. FITS files of LOFAR HBA can be found at \href{https://lofar-surveys.org/}{https://lofar-surveys.org/}.
Processed data used in this work will be made available upon reasonable request.


\bibliographystyle{mnras}
\bibliography{mybibliography}

\begin{thebibliography}{}
\makeatletter
\relax
\def\mn@urlcharsother{\let\do\@makeother \do\$\do\&\do\#\do\^\do\_\do\%\do\~}
\def\mn@doi{\begingroup\mn@urlcharsother \@ifnextchar [ {\mn@doi@} {\mn@doi@[]}}
\def\mn@doi@[#1]#2{\def\@tempa{#1}\ifx\@tempa\@empty \href {http://dx.doi.org/#2} {doi:#2}\else \href {http://dx.doi.org/#2} {#1}\fi \endgroup}
\def\mn@eprint#1#2{\mn@eprint@#1:#2::\@nil}
\def\mn@eprint@arXiv#1{\href {http://arxiv.org/abs/#1} {{\tt arXiv:#1}}}
\def\mn@eprint@dblp#1{\href {http://dblp.uni-trier.de/rec/bibtex/#1.xml} {dblp:#1}}
\def\mn@eprint@#1:#2:#3:#4\@nil{\def\@tempa {#1}\def\@tempb {#2}\def\@tempc {#3}\ifx \@tempc \@empty \let \@tempc \@tempb \let \@tempb \@tempa \fi \ifx \@tempb \@empty \def\@tempb {arXiv}\fi \@ifundefined {mn@eprint@\@tempb}{\@tempb:\@tempc}{\expandafter \expandafter \csname mn@eprint@\@tempb\endcsname \expandafter{\@tempc}}}

\bibitem[\protect\citeauthoryear{{Bonafede} et~al.,}{{Bonafede} et~al.}{2022}]{Bonafede2022}
{Bonafede} A.,  et~al., 2022, arXiv e-prints, \href {https://ui.adsabs.harvard.edu/abs/2022arXiv220301958B} {p. arXiv:2203.01958}

\bibitem[\protect\citeauthoryear{{Botteon} et~al.,}{{Botteon} et~al.}{2018}]{Botteon2018A1758}
{Botteon} A.,  et~al., 2018, \mn@doi [\mnras] {10.1093/mnras/sty1102}, \href {https://ui.adsabs.harvard.edu/abs/2018MNRAS.478..885B} {478, 885}

\bibitem[\protect\citeauthoryear{{Botteon} et~al.,}{{Botteon} et~al.}{2019}]{Botteon2019}
{Botteon} A.,  et~al., 2019, \mn@doi [\aap] {10.1051/0004-6361/201833861}, \href {https://ui.adsabs.harvard.edu/abs/2019A&A...622A..19B} {622, A19}

\bibitem[\protect\citeauthoryear{{Botteon} et~al.,}{{Botteon} et~al.}{2020}]{Botteon2020A1758}
{Botteon} A.,  et~al., 2020, \mn@doi [\mnras] {10.1093/mnrasl/slaa142}, \href {https://ui.adsabs.harvard.edu/abs/2020MNRAS.499L..11B} {499, L11}

\bibitem[\protect\citeauthoryear{{Botteon} et~al.,}{{Botteon} et~al.}{2021}]{Botteon2021}
{Botteon} A.,  et~al., 2021, \mn@doi [\aap] {10.1051/0004-6361/202040083}, \href {https://ui.adsabs.harvard.edu/abs/2021A&A...649A..37B} {649, A37}

\bibitem[\protect\citeauthoryear{{Botteon} et~al.,}{{Botteon} et~al.}{2022}]{Botteon2022DR2}
{Botteon} A.,  et~al., 2022, \mn@doi [\aap] {10.1051/0004-6361/202143020}, \href {https://ui.adsabs.harvard.edu/abs/2022A&A...660A..78B} {660, A78}

\bibitem[\protect\citeauthoryear{{Boxelaar}, {van Weeren}  \& {Botteon}}{{Boxelaar} et~al.}{2021}]{Boxelaar2021}
{Boxelaar} J.~M.,  {van Weeren} R.~J.,   {Botteon} A.,  2021, \mn@doi [Astronomy and Computing] {10.1016/j.ascom.2021.100464}, \href {https://ui.adsabs.harvard.edu/abs/2021A&C....3500464B} {35, 100464}

\bibitem[\protect\citeauthoryear{{Brunetti} \& {Jones}}{{Brunetti} \& {Jones}}{2014}]{Brunetti_and_Jones2014}
{Brunetti} G.,  {Jones} T.~W.,  2014, \mn@doi [International Journal of Modern Physics D] {10.1142/S0218271814300079}, \href {https://ui.adsabs.harvard.edu/abs/2014IJMPD..2330007B} {23, 1430007}

\bibitem[\protect\citeauthoryear{{Brunetti} et~al.,}{{Brunetti} et~al.}{2008}]{Brunetti2008}
{Brunetti} G.,  et~al., 2008, \mn@doi [\nat] {10.1038/nature07379}, \href {https://ui.adsabs.harvard.edu/abs/2008Natur.455..944B} {455, 944}

\bibitem[\protect\citeauthoryear{{Buote}}{{Buote}}{2001}]{Buote2001}
{Buote} D.~A.,  2001, \mn@doi [\apjl] {10.1086/320500}, \href {https://ui.adsabs.harvard.edu/abs/2001ApJ...553L..15B} {553, L15}

\bibitem[\protect\citeauthoryear{{Carilli}, {Perley}, {Dreher}  \& {Leahy}}{{Carilli} et~al.}{1991}]{Carilli1991}
{Carilli} C.~L.,  {Perley} R.~A.,  {Dreher} J.~W.,   {Leahy} J.~P.,  1991, \mn@doi [\apj] {10.1086/170813}, \href {https://ui.adsabs.harvard.edu/abs/1991ApJ...383..554C} {383, 554}

\bibitem[\protect\citeauthoryear{{Cassano} \& {Brunetti}}{{Cassano} \& {Brunetti}}{2005}]{Cassano_Brunetti_2005}
{Cassano} R.,  {Brunetti} G.,  2005, \mn@doi [\mnras] {10.1111/j.1365-2966.2005.08747.x}, \href {https://ui.adsabs.harvard.edu/abs/2005MNRAS.357.1313C} {357, 1313}

\bibitem[\protect\citeauthoryear{{Cassano}, {Brunetti}  \& {Setti}}{{Cassano} et~al.}{2006}]{Cassano2006}
{Cassano} R.,  {Brunetti} G.,   {Setti} G.,  2006, \mn@doi [\mnras] {10.1111/j.1365-2966.2006.10423.x}, \href {https://ui.adsabs.harvard.edu/abs/2006MNRAS.369.1577C} {369, 1577}

\bibitem[\protect\citeauthoryear{{Cassano}, {Ettori}, {Giacintucci}, {Brunetti}, {Markevitch}, {Venturi}  \& {Gitti}}{{Cassano} et~al.}{2010}]{Cassano2010}
{Cassano} R.,  {Ettori} S.,  {Giacintucci} S.,  {Brunetti} G.,  {Markevitch} M.,  {Venturi} T.,   {Gitti} M.,  2010, \mn@doi [\apjl] {10.1088/2041-8205/721/2/L82}, \href {https://ui.adsabs.harvard.edu/abs/2010ApJ...721L..82C} {721, L82}

\bibitem[\protect\citeauthoryear{{Cassano} et~al.,}{{Cassano} et~al.}{2013}]{Cassano2013}
{Cassano} R.,  et~al., 2013, \mn@doi [\apj] {10.1088/0004-637X/777/2/141}, \href {https://ui.adsabs.harvard.edu/abs/2013ApJ...777..141C} {777, 141}

\bibitem[\protect\citeauthoryear{{Cavaliere} \& {Fusco-Femiano}}{{Cavaliere} \& {Fusco-Femiano}}{1976}]{CavaliereFusco1976}
{Cavaliere} A.,  {Fusco-Femiano} R.,  1976, \aap, \href {https://ui.adsabs.harvard.edu/abs/1976A&A....49..137C} {49, 137}

\bibitem[\protect\citeauthoryear{{Chandra}, {Ray}  \& {Bhatnagar}}{{Chandra} et~al.}{2004}]{Chandra2004}
{Chandra} P.,  {Ray} A.,   {Bhatnagar} S.,  2004, \mn@doi [\apjl] {10.1086/383615}, \href {https://ui.adsabs.harvard.edu/abs/2004ApJ...604L..97C} {604, L97}

\bibitem[\protect\citeauthoryear{{Cuciti}, {Cassano}, {Brunetti}, {Dallacasa}, {Kale}, {Ettori}  \& {Venturi}}{{Cuciti} et~al.}{2015}]{Cuciti2015}
{Cuciti} V.,  {Cassano} R.,  {Brunetti} G.,  {Dallacasa} D.,  {Kale} R.,  {Ettori} S.,   {Venturi} T.,  2015, \mn@doi [\aap] {10.1051/0004-6361/201526420}, \href {https://ui.adsabs.harvard.edu/abs/2015A&A...580A..97C} {580, A97}

\bibitem[\protect\citeauthoryear{{Cuciti}, {Brunetti}, {van Weeren}, {Bonafede}, {Dallacasa}, {Cassano}, {Venturi}  \& {Kale}}{{Cuciti} et~al.}{2018}]{Cuciti2018}
{Cuciti} V.,  {Brunetti} G.,  {van Weeren} R.,  {Bonafede} A.,  {Dallacasa} D.,  {Cassano} R.,  {Venturi} T.,   {Kale} R.,  2018, \mn@doi [\aap] {10.1051/0004-6361/201731174}, \href {https://ui.adsabs.harvard.edu/abs/2018A&A...609A..61C} {609, A61}

\bibitem[\protect\citeauthoryear{{Cuciti} et~al.,}{{Cuciti} et~al.}{2021}]{Cuciti2021}
{Cuciti} V.,  et~al., 2021, \mn@doi [\aap] {10.1051/0004-6361/202039208}, \href {https://ui.adsabs.harvard.edu/abs/2021A&A...647A..51C} {647, A51}

\bibitem[\protect\citeauthoryear{{Cuciti} et~al.,}{{Cuciti} et~al.}{2022}]{Cuciti2022Nat}
{Cuciti} V.,  et~al., 2022, \mn@doi [\nat] {10.1038/s41586-022-05149-3}, \href {https://ui.adsabs.harvard.edu/abs/2022Natur.609..911C} {609, 911}

\bibitem[\protect\citeauthoryear{{Cuciti} et~al.,}{{Cuciti} et~al.}{2023}]{Cuciti2023}
{Cuciti} V.,  et~al., 2023, \mn@doi [\aap] {10.1051/0004-6361/202346755}, \href {https://ui.adsabs.harvard.edu/abs/2023A&A...680A..30C} {680, A30}

\bibitem[\protect\citeauthoryear{{Cutri} et~al.,}{{Cutri} et~al.}{2003}]{Cutri2003}
{Cutri} R.~M.,  et~al., 2003, {2MASS All Sky Catalog of point sources.}

\bibitem[\protect\citeauthoryear{{De Luca} \& {Molendi}}{{De Luca} \& {Molendi}}{2004}]{DeLucaMolendi}
{De Luca} {Molendi} 2004, \mn@doi [A&A] {10.1051/0004-6361:20034421}, 419, 837

\bibitem[\protect\citeauthoryear{{Duchesne}, {Johnston-Hollitt}  \& {Bartalucci}}{{Duchesne} et~al.}{2021}]{Duchesne2021}
{Duchesne} S.~W.,  {Johnston-Hollitt} M.,   {Bartalucci} I.,  2021, \mn@doi [\pasa] {10.1017/pasa.2021.45}, \href {https://ui.adsabs.harvard.edu/abs/2021PASA...38...53D} {38, e053}

\bibitem[\protect\citeauthoryear{{Eckert}, {Finoguenov}, {Ghirardini}, {Grandis}, {Kaefer}, {Sanders}  \& {Ramos-Ceja}}{{Eckert} et~al.}{2020}]{Eckert2020Opyprofit}
{Eckert} D.,  {Finoguenov} A.,  {Ghirardini} V.,  {Grandis} S.,  {Kaefer} F.,  {Sanders} J.,   {Ramos-Ceja} M.,  2020, \mn@doi [The Open Journal of Astrophysics] {10.21105/astro.2009.13944}, \href {https://ui.adsabs.harvard.edu/abs/2020OJAp....3E..12E} {3, 12}

\bibitem[\protect\citeauthoryear{{Edler} et~al.,}{{Edler} et~al.}{2022}]{Edler2022}
{Edler} H.~W.,  et~al., 2022, \mn@doi [\aap] {10.1051/0004-6361/202243737}, \href {https://ui.adsabs.harvard.edu/abs/2022A&A...666A...3E} {666, A3}

\bibitem[\protect\citeauthoryear{{En{\ss}lin} \& {Br{\"u}ggen}}{{En{\ss}lin} \& {Br{\"u}ggen}}{2002}]{EnsslinBrueggen2002}
{En{\ss}lin} T.~A.,  {Br{\"u}ggen} M.,  2002, \mn@doi [\mnras] {10.1046/j.1365-8711.2002.05261.x}, \href {https://ui.adsabs.harvard.edu/abs/2002MNRAS.331.1011E} {331, 1011}

\bibitem[\protect\citeauthoryear{{En{\ss}lin} \& {Gopal-Krishna}}{{En{\ss}lin} \& {Gopal-Krishna}}{2001}]{EnsslinKrishna2001}
{En{\ss}lin} T.~A.,  {Gopal-Krishna} 2001, \mn@doi [\aap] {10.1051/0004-6361:20000198}, \href {https://ui.adsabs.harvard.edu/abs/2001A&A...366...26E} {366, 26}

\bibitem[\protect\citeauthoryear{{Feretti} \& {Giovannini}}{{Feretti} \& {Giovannini}}{2008}]{FerettiGiovannini2008}
{Feretti} L.,  {Giovannini} G.,  2008, in {Plionis} M.,  {L{\'o}pez-Cruz} O.,   {Hughes} D.,  eds, , Vol.~740, A Pan-Chromatic View of Clusters of Galaxies and the Large-Scale Structure.
p.~24, \mn@doi{10.1007/978-1-4020-6941-3_5}

\bibitem[\protect\citeauthoryear{{Feretti}, {Giovannini}, {Klein}, {Mack}, {Sijbring}  \& {Zech}}{{Feretti} et~al.}{1998}]{Feretti1998}
{Feretti} L.,  {Giovannini} G.,  {Klein} U.,  {Mack} K.~H.,  {Sijbring} L.~G.,   {Zech} G.,  1998, \aap, \href {https://ui.adsabs.harvard.edu/abs/1998A&A...331..475F} {331, 475}

\bibitem[\protect\citeauthoryear{{Garon} et~al.,}{{Garon} et~al.}{2019}]{Garon2019}
{Garon} A.~F.,  et~al., 2019, \mn@doi [\aj] {10.3847/1538-3881/aaff62}, \href {https://ui.adsabs.harvard.edu/abs/2019AJ....157..126G} {157, 126}

\bibitem[\protect\citeauthoryear{{Giacintucci}, {Venturi}, {Murgia}, {Dallacasa}, {Athreya}, {Bardelli}, {Mazzotta}  \& {Saikia}}{{Giacintucci} et~al.}{2007}]{Giacintucci2007}
{Giacintucci} S.,  {Venturi} T.,  {Murgia} M.,  {Dallacasa} D.,  {Athreya} R.,  {Bardelli} S.,  {Mazzotta} P.,   {Saikia} D.~J.,  2007, \mn@doi [\aap] {10.1051/0004-6361:20077918}, \href {https://ui.adsabs.harvard.edu/abs/2007A&A...476...99G} {476, 99}

\bibitem[\protect\citeauthoryear{{Giacintucci} et~al.,}{{Giacintucci} et~al.}{2022}]{Giacintucci2022}
{Giacintucci} S.,  et~al., 2022, \mn@doi [\apj] {10.3847/1538-4357/ac7805}, \href {https://ui.adsabs.harvard.edu/abs/2022ApJ...934...49G} {934, 49}

\bibitem[\protect\citeauthoryear{{Hardcastle}}{{Hardcastle}}{2013}]{Hardcastle2013}
{Hardcastle} M.~J.,  2013, \mn@doi [\mnras] {10.1093/mnras/stt1024}, \href {https://ui.adsabs.harvard.edu/abs/2013MNRAS.433.3364H} {433, 3364}

\bibitem[\protect\citeauthoryear{{Hardcastle} \& {Croston}}{{Hardcastle} \& {Croston}}{2020}]{Hardcastle2020_review}
{Hardcastle} M.~J.,  {Croston} J.~H.,  2020, \mn@doi [\nar] {10.1016/j.newar.2020.101539}, \href {https://ui.adsabs.harvard.edu/abs/2020NewAR..8801539H} {88, 101539}

\bibitem[\protect\citeauthoryear{{Hardcastle}, {Birkinshaw}  \& {Worrall}}{{Hardcastle} et~al.}{1998}]{Hardcastle1998}
{Hardcastle} M.~J.,  {Birkinshaw} M.,   {Worrall} D.~M.,  1998, \mn@doi [\mnras] {10.1111/j.1365-8711.1998.01159.x}, \href {https://ui.adsabs.harvard.edu/abs/1998MNRAS.294..615H} {294, 615}

\bibitem[\protect\citeauthoryear{{Hardcastle} et~al.,}{{Hardcastle} et~al.}{2021}]{Hardcastle2021_HBAflux}
{Hardcastle} M.~J.,  et~al., 2021, \mn@doi [\aap] {10.1051/0004-6361/202038814}, \href {https://ui.adsabs.harvard.edu/abs/2021A&A...648A..10H} {648, A10}

\bibitem[\protect\citeauthoryear{{Harwood}, {Hardcastle}, {Croston}  \& {Goodger}}{{Harwood} et~al.}{2013}]{Harwood2013}
{Harwood} J.~J.,  {Hardcastle} M.~J.,  {Croston} J.~H.,   {Goodger} J.~L.,  2013, \mn@doi [\mnras] {10.1093/mnras/stt1526}, \href {https://ui.adsabs.harvard.edu/abs/2013MNRAS.435.3353H} {435, 3353}

\bibitem[\protect\citeauthoryear{{Hill} \& {Longair}}{{Hill} \& {Longair}}{1971}]{HillLongair1971}
{Hill} J.~M.,  {Longair} M.~S.,  1971, \mn@doi [\mnras] {10.1093/mnras/154.2.125}, \href {https://ui.adsabs.harvard.edu/abs/1971MNRAS.154..125H} {154, 125}

\bibitem[\protect\citeauthoryear{{Jaffe} \& {Perola}}{{Jaffe} \& {Perola}}{1973}]{JaffePerola1973}
{Jaffe} W.~J.,  {Perola} G.~C.,  1973, \aap, \href {https://ui.adsabs.harvard.edu/abs/1973A&A....26..423J} {26, 423}

\bibitem[\protect\citeauthoryear{{Jones}, {Nolting}, {O'Neill}  \& {Mendygral}}{{Jones} et~al.}{2017}]{Jones2017}
{Jones} T.~W.,  {Nolting} C.,  {O'Neill} B.~J.,   {Mendygral} P.~J.,  2017, \mn@doi [Physics of Plasmas] {10.1063/1.4978620}, \href {https://ui.adsabs.harvard.edu/abs/2017PhPl...24d1402J} {24, 041402}

\bibitem[\protect\citeauthoryear{{Kardashev}}{{Kardashev}}{1962}]{Kardashev1962}
{Kardashev} N.~S.,  1962, \sovast, \href {https://ui.adsabs.harvard.edu/abs/1962SvA.....6..317K} {6, 317}

\bibitem[\protect\citeauthoryear{{Katz-Stone}, {Rudnick}  \& {Anderson}}{{Katz-Stone} et~al.}{1993}]{Katz-Stone1993}
{Katz-Stone} D.~M.,  {Rudnick} L.,   {Anderson} M.~C.,  1993, \mn@doi [\apj] {10.1086/172536}, \href {https://ui.adsabs.harvard.edu/abs/1993ApJ...407..549K} {407, 549}

\bibitem[\protect\citeauthoryear{{Kempner}, {Blanton}, {Clarke}, {En{\ss}lin}, {Johnston-Hollitt}  \& {Rudnick}}{{Kempner} et~al.}{2004}]{Kempner2004}
{Kempner} J.~C.,  {Blanton} E.~L.,  {Clarke} T.~E.,  {En{\ss}lin} T.~A.,  {Johnston-Hollitt} M.,   {Rudnick} L.,  2004, in {Reiprich} T.,  {Kempner} J.,   {Soker} N.,  eds, The Riddle of Cooling Flows in Galaxies and Clusters of galaxies. p.~335 (\mn@eprint {arXiv} {astro-ph/0310263}), \mn@doi{10.48550/arXiv.astro-ph/0310263}

\bibitem[\protect\citeauthoryear{{Klamer}, {Subrahmanyan}  \& {Hunstead}}{{Klamer} et~al.}{2004}]{Klamer2004}
{Klamer} I.,  {Subrahmanyan} R.,   {Hunstead} R.~W.,  2004, \mn@doi [\mnras] {10.1111/j.1365-2966.2004.07757.x}, \href {https://ui.adsabs.harvard.edu/abs/2004MNRAS.351..101K} {351, 101}

\bibitem[\protect\citeauthoryear{{Kundu}, {Vaidya}  \& {Mignone}}{{Kundu} et~al.}{2021}]{Kundu2021}
{Kundu} S.,  {Vaidya} B.,   {Mignone} A.,  2021, \mn@doi [\apj] {10.3847/1538-4357/ac1ba5}, \href {https://ui.adsabs.harvard.edu/abs/2021ApJ...921...74K} {921, 74}

\bibitem[\protect\citeauthoryear{{Kuntz} \& {Snowden}}{{Kuntz} \& {Snowden}}{2008}]{KuntzSnowden}
{Kuntz} {Snowden} 2008, \mn@doi [A&A] {10.1051/0004-6361:20077912}, 478, 575

\bibitem[\protect\citeauthoryear{{Lal}}{{Lal}}{2020}]{Lal2020}
{Lal} D.~V.,  2020, \mn@doi [\aj] {10.3847/1538-3881/abacd1}, \href {https://ui.adsabs.harvard.edu/abs/2020AJ....160..161L} {160, 161}

\bibitem[\protect\citeauthoryear{{Leccardi} \& {Molendi}}{{Leccardi} \& {Molendi}}{2008}]{Leccardi}
{Leccardi} A.,  {Molendi} S.,  2008, \mn@doi [\aap] {10.1051/0004-6361:200809538}, \href {https://ui.adsabs.harvard.edu/abs/2008A&A...486..359L} {486, 359}

\bibitem[\protect\citeauthoryear{{Longair}}{{Longair}}{2011}]{Longair2011book}
{Longair} M.~S.,  2011, {High Energy Astrophysics}

\bibitem[\protect\citeauthoryear{{Lovisari} et~al.,}{{Lovisari} et~al.}{2017}]{Lovisari2017}
{Lovisari} L.,  et~al., 2017, \mn@doi [\apj] {10.3847/1538-4357/aa855f}, \href {https://ui.adsabs.harvard.edu/abs/2017ApJ...846...51L} {846, 51}

\bibitem[\protect\citeauthoryear{{Mandal} et~al.,}{{Mandal} et~al.}{2019}]{Mandal2019}
{Mandal} S.,  et~al., 2019, \mn@doi [\aap] {10.1051/0004-6361/201833992}, \href {https://ui.adsabs.harvard.edu/abs/2019A&A...622A..22M} {622, A22}

\bibitem[\protect\citeauthoryear{{Mandal} et~al.,}{{Mandal} et~al.}{2020}]{Mandal2020}
{Mandal} S.,  et~al., 2020, \mn@doi [\aap] {10.1051/0004-6361/201936560}, \href {https://ui.adsabs.harvard.edu/abs/2020A&A...634A...4M} {634, A4}

\bibitem[\protect\citeauthoryear{{Mao}, {Sharp}, {Saikia}, {Norris}, {Johnston-Hollitt}, {Middelberg}  \& {Lovell}}{{Mao} et~al.}{2010}]{Mao2010}
{Mao} M.~Y.,  {Sharp} R.,  {Saikia} D.~J.,  {Norris} R.~P.,  {Johnston-Hollitt} M.,  {Middelberg} E.,   {Lovell} J. E.~J.,  2010, \mn@doi [\mnras] {10.1111/j.1365-2966.2010.16853.x}, \href {https://ui.adsabs.harvard.edu/abs/2010MNRAS.406.2578M} {406, 2578}

\bibitem[\protect\citeauthoryear{{McMullin}, {Waters}, {Schiebel}, {Young}  \& {Golap}}{{McMullin} et~al.}{2007}]{McMullin2007_CASA}
{McMullin} J.~P.,  {Waters} B.,  {Schiebel} D.,  {Young} W.,   {Golap} K.,  2007, in {Shaw} R.~A.,  {Hill} F.,   {Bell} D.~J.,  eds,  Astronomical Society of the Pacific Conference Series Vol. 376, Astronomical Data Analysis Software and Systems XVI. p.~127

\bibitem[\protect\citeauthoryear{{Migkas}, {Schellenberger}, {Reiprich}, {Pacaud}, {Ramos-Ceja}  \& {Lovisari}}{{Migkas} et~al.}{2020}]{Migkas2020}
{Migkas} K.,  {Schellenberger} G.,  {Reiprich} T.~H.,  {Pacaud} F.,  {Ramos-Ceja} M.~E.,   {Lovisari} L.,  2020, \mn@doi [\aap] {10.1051/0004-6361/201936602}, \href {https://ui.adsabs.harvard.edu/abs/2020A&A...636A..15M} {636, A15}

\bibitem[\protect\citeauthoryear{{Mignone}, {Massaglia}  \& {Bodo}}{{Mignone} et~al.}{2004}]{Mignone2004}
{Mignone} A.,  {Massaglia} S.,   {Bodo} G.,  2004, \mn@doi [\apss] {10.1023/B:ASTR.0000044668.90635.f7}, \href {https://ui.adsabs.harvard.edu/abs/2004Ap&SS.293..199M} {293, 199}

\bibitem[\protect\citeauthoryear{{Miley}}{{Miley}}{1973}]{Miley1973}
{Miley} G.~K.,  1973, \aap, \href {https://ui.adsabs.harvard.edu/abs/1973A&A....26..413M} {26, 413}

\bibitem[\protect\citeauthoryear{{Miley}}{{Miley}}{1980}]{Miley1980}
{Miley} G.,  1980, \mn@doi [\araa] {10.1146/annurev.aa.18.090180.001121}, \href {https://ui.adsabs.harvard.edu/abs/1980ARA&A..18..165M} {18, 165}

\bibitem[\protect\citeauthoryear{{Mukherjee}, {Bodo}, {Rossi}, {Mignone}  \& {Vaidya}}{{Mukherjee} et~al.}{2021}]{Mukherjee2021}
{Mukherjee} D.,  {Bodo} G.,  {Rossi} P.,  {Mignone} A.,   {Vaidya} B.,  2021, \mn@doi [\mnras] {10.1093/mnras/stab1327}, \href {https://ui.adsabs.harvard.edu/abs/2021MNRAS.505.2267M} {505, 2267}

\bibitem[\protect\citeauthoryear{{M{\"u}ller} et~al.,}{{M{\"u}ller} et~al.}{2021}]{Muller2021}
{M{\"u}ller} A.,  et~al., 2021, \mn@doi [\mnras] {10.1093/mnras/stab2928}, \href {https://ui.adsabs.harvard.edu/abs/2021MNRAS.508.5326M} {508, 5326}

\bibitem[\protect\citeauthoryear{{Murgia}, {Govoni}, {Markevitch}, {Feretti}, {Giovannini}, {Taylor}  \& {Carretti}}{{Murgia} et~al.}{2009}]{Murgia2009}
{Murgia} M.,  {Govoni} F.,  {Markevitch} M.,  {Feretti} L.,  {Giovannini} G.,  {Taylor} G.~B.,   {Carretti} E.,  2009, \mn@doi [\aap] {10.1051/0004-6361/200911659}, \href {https://ui.adsabs.harvard.edu/abs/2009A&A...499..679M} {499, 679}

\bibitem[\protect\citeauthoryear{{Nasa High Energy Astrophysics Science Archive Research Center (Heasarc)}}{{Nasa High Energy Astrophysics Science Archive Research Center (Heasarc)}}{2014}]{HEAsoft2014}
{Nasa High Energy Astrophysics Science Archive Research Center (Heasarc)} 2014, {HEAsoft: Unified Release of FTOOLS and XANADU}, Astrophysics Source Code Library, record ascl:1408.004 (\mn@eprint {ascl} {1408.004})

\bibitem[\protect\citeauthoryear{{Nolting}, {Jones}, {O'Neill}  \& {Mendygral}}{{Nolting} et~al.}{2019a}]{Nolting2019_876Aligned}
{Nolting} C.,  {Jones} T.~W.,  {O'Neill} B.~J.,   {Mendygral} P.~J.,  2019a, \mn@doi [\apj] {10.3847/1538-4357/ab16d6}, \href {https://ui.adsabs.harvard.edu/abs/2019ApJ...876..154N} {876, 154}

\bibitem[\protect\citeauthoryear{{Nolting}, {Jones}, {O'Neill}  \& {Mendygral}}{{Nolting} et~al.}{2019b}]{Nolting2019_885Orthogonal}
{Nolting} C.,  {Jones} T.~W.,  {O'Neill} B.~J.,   {Mendygral} P.~J.,  2019b, \mn@doi [\apj] {10.3847/1538-4357/ab4650}, \href {https://ui.adsabs.harvard.edu/abs/2019ApJ...885...80N} {885, 80}

\bibitem[\protect\citeauthoryear{{O'Dea} \& {Owen}}{{O'Dea} \& {Owen}}{1985}]{ODeaOwen1985}
{O'Dea} C.~P.,  {Owen} F.~N.,  1985, \mn@doi [\aj] {10.1086/113801}, \href {https://ui.adsabs.harvard.edu/abs/1985AJ.....90..927O} {90, 927}

\bibitem[\protect\citeauthoryear{{O'Neill}, {Jones}, {Nolting}  \& {Mendygral}}{{O'Neill} et~al.}{2019a}]{O'Neill2019_884}
{O'Neill} B.~J.,  {Jones} T.~W.,  {Nolting} C.,   {Mendygral} P.~J.,  2019a, \mn@doi [\apj] {10.3847/1538-4357/ab40b1}, \href {https://ui.adsabs.harvard.edu/abs/2019ApJ...884...12O} {884, 12}

\bibitem[\protect\citeauthoryear{{O'Neill}, {Jones}, {Nolting}  \& {Mendygral}}{{O'Neill} et~al.}{2019b}]{ONeill2019_887}
{O'Neill} B.~J.,  {Jones} T.~W.,  {Nolting} C.,   {Mendygral} P.~J.,  2019b, \mn@doi [\apj] {10.3847/1538-4357/ab4efa}, \href {https://ui.adsabs.harvard.edu/abs/2019ApJ...887...26O} {887, 26}

\bibitem[\protect\citeauthoryear{{Offringa} et~al.,}{{Offringa} et~al.}{2014}]{Offringa2014}
{Offringa} A.~R.,  et~al., 2014, \mn@doi [\mnras] {10.1093/mnras/stu1368}, \href {https://ui.adsabs.harvard.edu/abs/2014MNRAS.444..606O} {444, 606}

\bibitem[\protect\citeauthoryear{{Ohmura}, {Asano}, {Nishiwaki}, {Machida}  \& {Sakemi}}{{Ohmura} et~al.}{2023}]{Ohmura2023}
{Ohmura} T.,  {Asano} K.,  {Nishiwaki} K.,  {Machida} M.,   {Sakemi} H.,  2023, \mn@doi [\apj] {10.3847/1538-4357/acd338}, \href {https://ui.adsabs.harvard.edu/abs/2023ApJ...951...76O} {951, 76}

\bibitem[\protect\citeauthoryear{{Owen} \& {Rudnick}}{{Owen} \& {Rudnick}}{1976}]{OwenRudnick1976}
{Owen} F.~N.,  {Rudnick} L.,  1976, \mn@doi [\apjl] {10.1086/182077}, \href {https://ui.adsabs.harvard.edu/abs/1976ApJ...205L...1O} {205, L1}

\bibitem[\protect\citeauthoryear{Pacaud et~al.,}{Pacaud et~al.}{2006}]{Pacaud2006}
Pacaud F.,  et~al., 2006, \mn@doi [Monthly Notices of the Royal Astronomical Society] {10.1111/j.1365-2966.2006.10881.x}, 372, 578–590

\bibitem[\protect\citeauthoryear{{Pacholczyk}}{{Pacholczyk}}{1970}]{Pacholczyk1970book}
{Pacholczyk} A.~G.,  1970, {Radio astrophysics. Nonthermal processes in galactic and extragalactic sources}

\bibitem[\protect\citeauthoryear{{Pal} \& {Kumari}}{{Pal} \& {Kumari}}{2023}]{Pal2023}
{Pal} S.,  {Kumari} S.,  2023, \mn@doi [Journal of Astrophysics and Astronomy] {10.1007/s12036-022-09892-x}, \href {https://ui.adsabs.harvard.edu/abs/2023JApA...44...17P} {44, 17}

\bibitem[\protect\citeauthoryear{{P{\^a}ris} et~al.,}{{P{\^a}ris} et~al.}{2014}]{Paris2014_SDSS-DR10}
{P{\^a}ris} I.,  et~al., 2014, \mn@doi [\aap] {10.1051/0004-6361/201322691}, \href {https://ui.adsabs.harvard.edu/abs/2014A&A...563A..54P} {563, A54}

\bibitem[\protect\citeauthoryear{{Parma}, {Murgia}, {Morganti}, {Capetti}, {de Ruiter}  \& {Fanti}}{{Parma} et~al.}{1999}]{Parma1999}
{Parma} P.,  {Murgia} M.,  {Morganti} R.,  {Capetti} A.,  {de Ruiter} H.~R.,   {Fanti} R.,  1999, \mn@doi [\aap] {10.48550/arXiv.astro-ph/9812413}, \href {https://ui.adsabs.harvard.edu/abs/1999A&A...344....7P} {344, 7}

\bibitem[\protect\citeauthoryear{{Pasini} et~al.,}{{Pasini} et~al.}{2022}]{Pasini2022}
{Pasini} T.,  et~al., 2022, \mn@doi [\aap] {10.1051/0004-6361/202243833}, \href {https://ui.adsabs.harvard.edu/abs/2022A&A...663A.105P} {663, A105}

\bibitem[\protect\citeauthoryear{{Perley} \& {Butler}}{{Perley} \& {Butler}}{2013}]{Perley2013}
{Perley} R.~A.,  {Butler} B.~J.,  2013, \mn@doi [\apjs] {10.1088/0067-0049/204/2/19}, \href {https://ui.adsabs.harvard.edu/abs/2013ApJS..204...19P} {204, 19}

\bibitem[\protect\citeauthoryear{{Pfrommer} \& {Jones}}{{Pfrommer} \& {Jones}}{2011}]{PfrommerJones2011}
{Pfrommer} C.,  {Jones} T.~W.,  2011, \mn@doi [\apj] {10.1088/0004-637X/730/1/22}, \href {https://ui.adsabs.harvard.edu/abs/2011ApJ...730...22P} {730, 22}

\bibitem[\protect\citeauthoryear{{Planck Collaboration} et~al.,}{{Planck Collaboration} et~al.}{2016}]{PSZ2_2016}
{Planck Collaboration} et~al., 2016, \mn@doi [\aap] {10.1051/0004-6361/201525823}, \href {https://ui.adsabs.harvard.edu/abs/2016A&A...594A..27P} {594, A27}

\bibitem[\protect\citeauthoryear{{Ramos-Ceja}, {Pacaud, F.}, {Reiprich, T. H.}, {Migkas, K.}, {Lovisari, L.}  \& {Schellenberger, G.}}{{Ramos-Ceja} et~al.}{2019}]{FWC}
{Ramos-Ceja} {Pacaud, F.} {Reiprich, T. H.} {Migkas, K.} {Lovisari, L.}  {Schellenberger, G.} 2019, \mn@doi [A&A] {10.1051/0004-6361/201935111}, 626, A48

\bibitem[\protect\citeauthoryear{{Rossetti}, {Gastaldello}, {Eckert}, {Della Torre}, {Pantiri}, {Cazzoletti}  \& {Molendi}}{{Rossetti} et~al.}{2017}]{Rossetti2017_ref_redshift}
{Rossetti} M.,  {Gastaldello} F.,  {Eckert} D.,  {Della Torre} M.,  {Pantiri} G.,  {Cazzoletti} P.,   {Molendi} S.,  2017, \mn@doi [\mnras] {10.1093/mnras/stx493}, \href {https://ui.adsabs.harvard.edu/abs/2017MNRAS.468.1917R} {468, 1917}

\bibitem[\protect\citeauthoryear{{Rudnick} et~al.,}{{Rudnick} et~al.}{2022}]{Rudnick2022}
{Rudnick} L.,  et~al., 2022, \mn@doi [\apj] {10.3847/1538-4357/ac7c76}, \href {https://ui.adsabs.harvard.edu/abs/2022ApJ...935..168R} {935, 168}

\bibitem[\protect\citeauthoryear{{Ryle} \& {Windram}}{{Ryle} \& {Windram}}{1968}]{RyleWindram1968}
{Ryle} M.,  {Windram} M.~D.,  1968, \mn@doi [\mnras] {10.1093/mnras/138.1.1}, \href {https://ui.adsabs.harvard.edu/abs/1968MNRAS.138....1R} {138, 1}

\bibitem[\protect\citeauthoryear{{Sanders} et~al.,}{{Sanders} et~al.}{2016a}]{Sanders2016MNRAS457}
{Sanders} J.~S.,  et~al., 2016a, \mn@doi [\mnras] {10.1093/mnras/stv2972}, \href {https://ui.adsabs.harvard.edu/abs/2016MNRAS.457...82S} {457, 82}

\bibitem[\protect\citeauthoryear{{Sanders}, {Fabian}, {Russell}, {Walker}  \& {Blundell}}{{Sanders} et~al.}{2016b}]{Sanders2016MNRAS460}
{Sanders} J.~S.,  {Fabian} A.~C.,  {Russell} H.~R.,  {Walker} S.~A.,   {Blundell} K.~M.,  2016b, \mn@doi [\mnras] {10.1093/mnras/stw1119}, \href {https://ui.adsabs.harvard.edu/abs/2016MNRAS.460.1898S} {460, 1898}

\bibitem[\protect\citeauthoryear{{Sanders}, {Walker}, {ZuHone}  \& {Bellomi}}{{Sanders} et~al.}{2018}]{Sanders2018ChandraNews}
{Sanders} J.,  {Walker} S.,  {ZuHone} J.,   {Bellomi} E.,  2018, Chandra News, \href {https://ui.adsabs.harvard.edu/abs/2018ChNew..25....1S} {25, 1}

\bibitem[\protect\citeauthoryear{{Sanders} et~al.,}{{Sanders} et~al.}{2022}]{Sanders2022}
{Sanders} J.~S.,  et~al., 2022, \mn@doi [\aap] {10.1051/0004-6361/202141501}, \href {https://ui.adsabs.harvard.edu/abs/2022A&A...661A..36S} {661, A36}

\bibitem[\protect\citeauthoryear{{Sebastian}, {Lal}  \& {Pramesh Rao}}{{Sebastian} et~al.}{2017}]{Sebastian2017}
{Sebastian} B.,  {Lal} D.~V.,   {Pramesh Rao} A.,  2017, \mn@doi [\aj] {10.3847/1538-3881/aa88d0}, \href {https://ui.adsabs.harvard.edu/abs/2017AJ....154..169S} {154, 169}

\bibitem[\protect\citeauthoryear{{Shimwell} et~al.,}{{Shimwell} et~al.}{2017}]{Shimwell2017}
{Shimwell} T.~W.,  et~al., 2017, \mn@doi [\aap] {10.1051/0004-6361/201629313}, \href {https://ui.adsabs.harvard.edu/abs/2017A&A...598A.104S} {598, A104}

\bibitem[\protect\citeauthoryear{{Shimwell} et~al.,}{{Shimwell} et~al.}{2019}]{Shimwell2019}
{Shimwell} T.~W.,  et~al., 2019, \mn@doi [\aap] {10.1051/0004-6361/201833559}, \href {https://ui.adsabs.harvard.edu/abs/2019A&A...622A...1S} {622, A1}

\bibitem[\protect\citeauthoryear{{Shimwell} et~al.,}{{Shimwell} et~al.}{2022}]{Shimwell2022}
{Shimwell} T.~W.,  et~al., 2022, \mn@doi [\aap] {10.1051/0004-6361/202142484}, \href {https://ui.adsabs.harvard.edu/abs/2022A&A...659A...1S} {659, A1}

\bibitem[\protect\citeauthoryear{{Sijbring} \& {de Bruyn}}{{Sijbring} \& {de Bruyn}}{1998}]{Sijbring1998}
{Sijbring} D.,  {de Bruyn} A.~G.,  1998, \aap, \href {https://ui.adsabs.harvard.edu/abs/1998A&A...331..901S} {331, 901}

\bibitem[\protect\citeauthoryear{{Smirnov} \& {Tasse}}{{Smirnov} \& {Tasse}}{2015}]{Smirnov2015}
{Smirnov} O.~M.,  {Tasse} C.,  2015, \mn@doi [\mnras] {10.1093/mnras/stv418}, \href {https://ui.adsabs.harvard.edu/abs/2015MNRAS.449.2668S} {449, 2668}

\bibitem[\protect\citeauthoryear{{Srivastava} \& {Singal}}{{Srivastava} \& {Singal}}{2020}]{Srivastava2020}
{Srivastava} S.,  {Singal} A.~K.,  2020, \mn@doi [\mnras] {10.1093/mnras/staa520}, \href {https://ui.adsabs.harvard.edu/abs/2020MNRAS.493.3811S} {493, 3811}

\bibitem[\protect\citeauthoryear{{Stroe}, {van Weeren}, {Intema}, {R{\"o}ttgering}, {Br{\"u}ggen}  \& {Hoeft}}{{Stroe} et~al.}{2013}]{Stroe2013}
{Stroe} A.,  {van Weeren} R.~J.,  {Intema} H.~T.,  {R{\"o}ttgering} H.~J.~A.,  {Br{\"u}ggen} M.,   {Hoeft} M.,  2013, \mn@doi [\aap] {10.1051/0004-6361/201321267}, \href {https://ui.adsabs.harvard.edu/abs/2013A&A...555A.110S} {555, A110}

\bibitem[\protect\citeauthoryear{{Tasse} et~al.,}{{Tasse} et~al.}{2018}]{Tasse2018}
{Tasse} C.,  et~al., 2018, \mn@doi [\aap] {10.1051/0004-6361/201731474}, \href {https://ui.adsabs.harvard.edu/abs/2018A&A...611A..87T} {611, A87}

\bibitem[\protect\citeauthoryear{{Tasse} et~al.,}{{Tasse} et~al.}{2021}]{Tasse2021}
{Tasse} C.,  et~al., 2021, \mn@doi [\aap] {10.1051/0004-6361/202038804}, \href {https://ui.adsabs.harvard.edu/abs/2021A&A...648A...1T} {648, A1}

\bibitem[\protect\citeauthoryear{{Tribble}}{{Tribble}}{1993}]{Tribble1993}
{Tribble} P.~C.,  1993, \mn@doi [\mnras] {10.1093/mnras/261.1.57}, \href {https://ui.adsabs.harvard.edu/abs/1993MNRAS.261...57T} {261, 57}

\bibitem[\protect\citeauthoryear{{Venkatesan}, {Batuski}, {Hanisch}  \& {Burns}}{{Venkatesan} et~al.}{1994}]{Venkatesan1994}
{Venkatesan} T.~C.~A.,  {Batuski} D.~J.,  {Hanisch} R.~J.,   {Burns} J.~O.,  1994, \mn@doi [\apj] {10.1086/174881}, \href {https://ui.adsabs.harvard.edu/abs/1994ApJ...436...67V} {436, 67}

\bibitem[\protect\citeauthoryear{Veronica et~al.,}{Veronica et~al.}{2022}]{Veronica_2022}
Veronica A.,  et~al., 2022, \mn@doi [Astronomy & Astrophysics] {10.1051/0004-6361/202141415}, 661, A46

\bibitem[\protect\citeauthoryear{{Wilber} et~al.,}{{Wilber} et~al.}{2018}]{Wilber2018}
{Wilber} A.,  et~al., 2018, \mn@doi [\mnras] {10.1093/mnras/stx2568}, \href {https://ui.adsabs.harvard.edu/abs/2018MNRAS.473.3536W} {473, 3536}

\bibitem[\protect\citeauthoryear{{Wilber} et~al.,}{{Wilber} et~al.}{2019}]{Wilber2019}
{Wilber} A.,  et~al., 2019, \mn@doi [\aap] {10.1051/0004-6361/201833884}, \href {https://ui.adsabs.harvard.edu/abs/2019A&A...622A..25W} {622, A25}

\bibitem[\protect\citeauthoryear{{Zhou} et~al.,}{{Zhou} et~al.}{2021}]{Zhou2021}
{Zhou} R.,  et~al., 2021, \mn@doi [\mnras] {10.1093/mnras/staa3764}, \href {https://ui.adsabs.harvard.edu/abs/2021MNRAS.501.3309Z} {501, 3309}

\bibitem[\protect\citeauthoryear{{de Gasperin}}{{de Gasperin}}{2017}]{deGasperin2017MNRAS}
{de Gasperin} F.,  2017, \mn@doi [\mnras] {10.1093/mnras/stx210}, \href {https://ui.adsabs.harvard.edu/abs/2017MNRAS.467.2234D} {467, 2234}

\bibitem[\protect\citeauthoryear{{de Gasperin}, {Ogrean}, {van Weeren}, {Dawson}, {Br{\"u}ggen}, {Bonafede}  \& {Simionescu}}{{de Gasperin} et~al.}{2015}]{deGasperin2015}
{de Gasperin} F.,  {Ogrean} G.~A.,  {van Weeren} R.~J.,  {Dawson} W.~A.,  {Br{\"u}ggen} M.,  {Bonafede} A.,   {Simionescu} A.,  2015, \mn@doi [\mnras] {10.1093/mnras/stv129}, \href {https://ui.adsabs.harvard.edu/abs/2015MNRAS.448.2197D} {448, 2197}

\bibitem[\protect\citeauthoryear{{de Gasperin} et~al.,}{{de Gasperin} et~al.}{2019}]{deGasperin2019}
{de Gasperin} F.,  et~al., 2019, \mn@doi [\aap] {10.1051/0004-6361/201833867}, \href {https://ui.adsabs.harvard.edu/abs/2019A&A...622A...5D} {622, A5}

\bibitem[\protect\citeauthoryear{{de Gasperin} et~al.,}{{de Gasperin} et~al.}{2020}]{deGasperin2020}
{de Gasperin} F.,  et~al., 2020, \mn@doi [\aap] {10.1051/0004-6361/202038663}, \href {https://ui.adsabs.harvard.edu/abs/2020A&A...642A..85D} {642, A85}

\bibitem[\protect\citeauthoryear{{de Gasperin} et~al.,}{{de Gasperin} et~al.}{2021}]{deGasperin2021}
{de Gasperin} F.,  et~al., 2021, \mn@doi [\aap] {10.1051/0004-6361/202140316}, \href {https://ui.adsabs.harvard.edu/abs/2021A&A...648A.104D} {648, A104}

\bibitem[\protect\citeauthoryear{{de Gasperin} et~al.,}{{de Gasperin} et~al.}{2023}]{deGasperin2023}
{de Gasperin} F.,  et~al., 2023, \mn@doi [\aap] {10.1051/0004-6361/202245389}, \href {https://ui.adsabs.harvard.edu/abs/2023A&A...673A.165D} {673, A165}

\bibitem[\protect\citeauthoryear{{van Haarlem} et~al.,}{{van Haarlem} et~al.}{2013a}]{vanHaarlem2013_LOFAR}
{van Haarlem} M.~P.,  et~al., 2013a, \mn@doi [\aap] {10.1051/0004-6361/201220873}, \href {https://ui.adsabs.harvard.edu/abs/2013A&A...556A...2V} {556, A2}

\bibitem[\protect\citeauthoryear{{van Haarlem} et~al.,}{{van Haarlem} et~al.}{2013b}]{vanHaarlem2013}
{van Haarlem} M.~P.,  et~al., 2013b, \mn@doi [\aap] {10.1051/0004-6361/201220873}, \href {https://ui.adsabs.harvard.edu/abs/2013A&A...556A...2V} {556, A2}

\bibitem[\protect\citeauthoryear{{van Weeren}, {R{\"o}ttgering}, {Br{\"u}ggen}  \& {Cohen}}{{van Weeren} et~al.}{2009}]{vanWeeren2009}
{van Weeren} R.~J.,  {R{\"o}ttgering} H.~J.~A.,  {Br{\"u}ggen} M.,   {Cohen} A.,  2009, \mn@doi [\aap] {10.1051/0004-6361/200912501}, \href {https://ui.adsabs.harvard.edu/abs/2009A&A...508...75V} {508, 75}

\bibitem[\protect\citeauthoryear{{van Weeren}, {R{\"o}ttgering}  \& {Br{\"u}ggen}}{{van Weeren} et~al.}{2011}]{vanWeeren2011}
{van Weeren} R.~J.,  {R{\"o}ttgering} H.~J.~A.,   {Br{\"u}ggen} M.,  2011, \mn@doi [\aap] {10.1051/0004-6361/201015991}, \href {https://ui.adsabs.harvard.edu/abs/2011A&A...527A.114V} {527, A114}

\bibitem[\protect\citeauthoryear{{van Weeren} et~al.,}{{van Weeren} et~al.}{2017}]{vanWeeren2017NatAs}
{van Weeren} R.~J.,  et~al., 2017, \mn@doi [Nature Astronomy] {10.1038/s41550-016-0005}, \href {https://ui.adsabs.harvard.edu/abs/2017NatAs...1E...5V} {1, 0005}

\bibitem[\protect\citeauthoryear{{van Weeren}, {de Gasperin}, {Akamatsu}, {Br{\"u}ggen}, {Feretti}, {Kang}, {Stroe}  \& {Zandanel}}{{van Weeren} et~al.}{2019}]{vanWeeren2019}
{van Weeren} R.~J.,  {de Gasperin} F.,  {Akamatsu} H.,  {Br{\"u}ggen} M.,  {Feretti} L.,  {Kang} H.,  {Stroe} A.,   {Zandanel} F.,  2019, \mn@doi [\ssr] {10.1007/s11214-019-0584-z}, \href {https://ui.adsabs.harvard.edu/abs/2019SSRv..215...16V} {215, 16}

\bibitem[\protect\citeauthoryear{{van Weeren} et~al.,}{{van Weeren} et~al.}{2021}]{vanWeeren2021extraction}
{van Weeren} R.~J.,  et~al., 2021, \mn@doi [\aap] {10.1051/0004-6361/202039826}, \href {https://ui.adsabs.harvard.edu/abs/2021A&A...651A.115V} {651, A115}

\makeatother
\end{thebibliography}

\onecolumn
\appendix

\newpage
\section{Flux Density CAlculator results}
\label{appendix:FDCA}

\reviewfirst{
Flux Density CAlculator (\texttt{FDCA}) results for the 2D surface brightness fitting to study the central radio halo (see \cref{subsec:radio_halo}). The first column refers to the original source-subtracted images shown in \cref{fig:halos_allfreq}.
}
\vspace{-3cm}
\begin{figure*}
     \begin{subfigure}[b]{\textwidth}
         \centering
         \includegraphics[width=\textwidth]{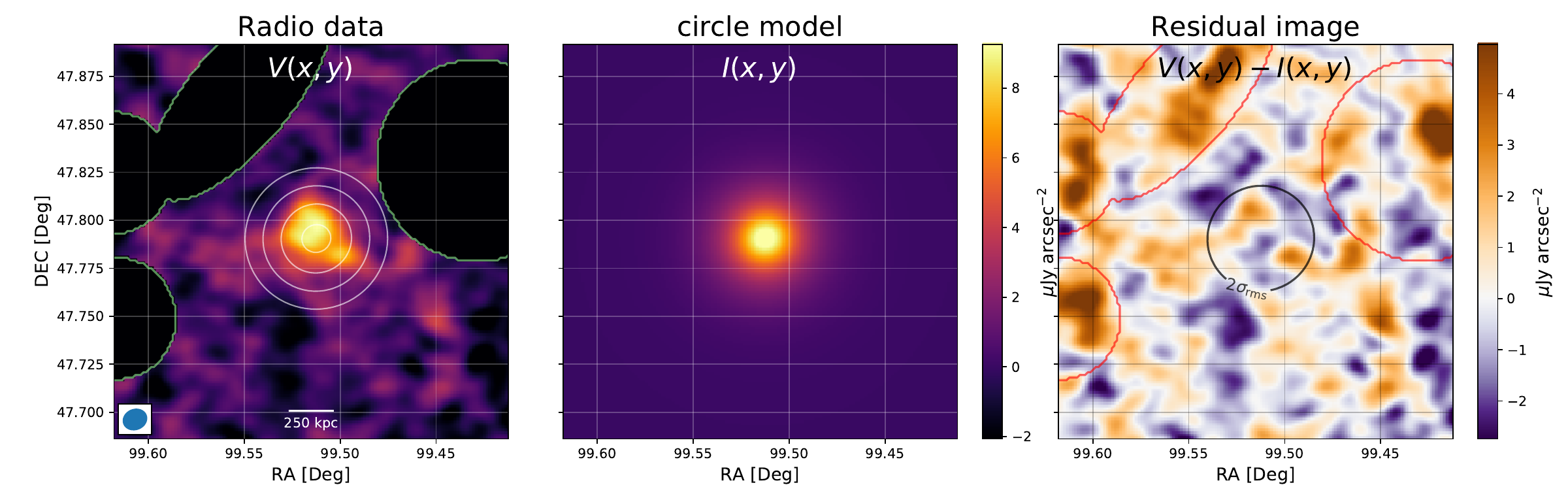}
     \end{subfigure}
    \vspace{-0.5cm}
     \begin{subfigure}[b]{\textwidth}
         \centering
         \includegraphics[width=\textwidth]{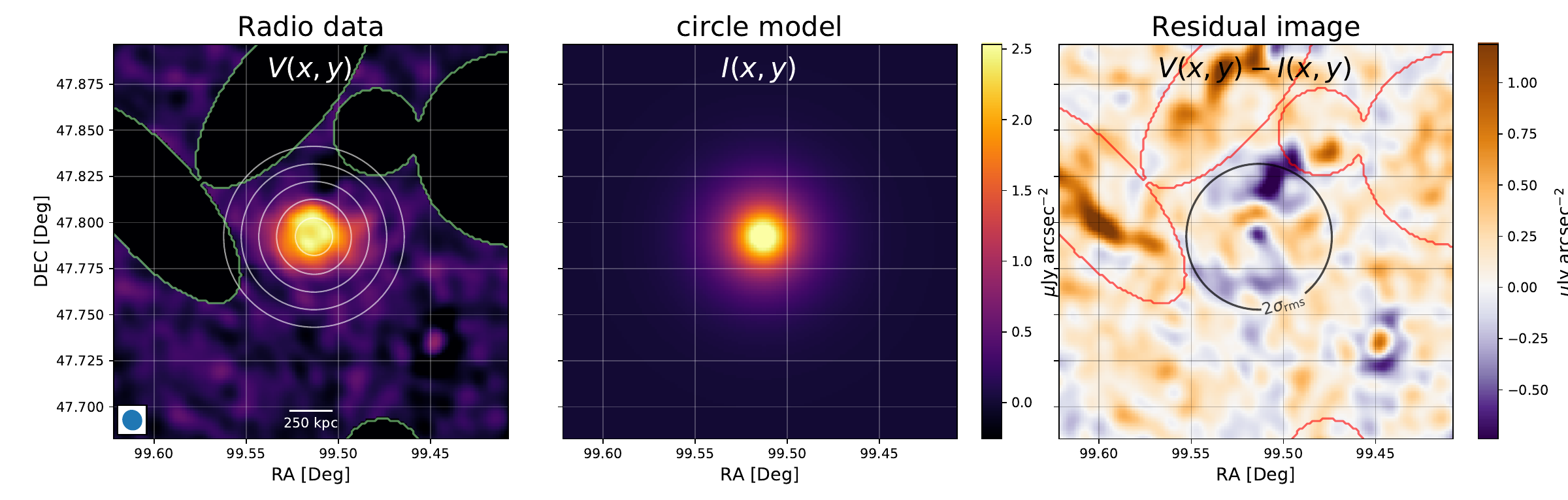}
     \end{subfigure}
    \vspace{-0.5cm}
     \begin{subfigure}[b]{\textwidth}
         \centering
         \includegraphics[width=\textwidth]{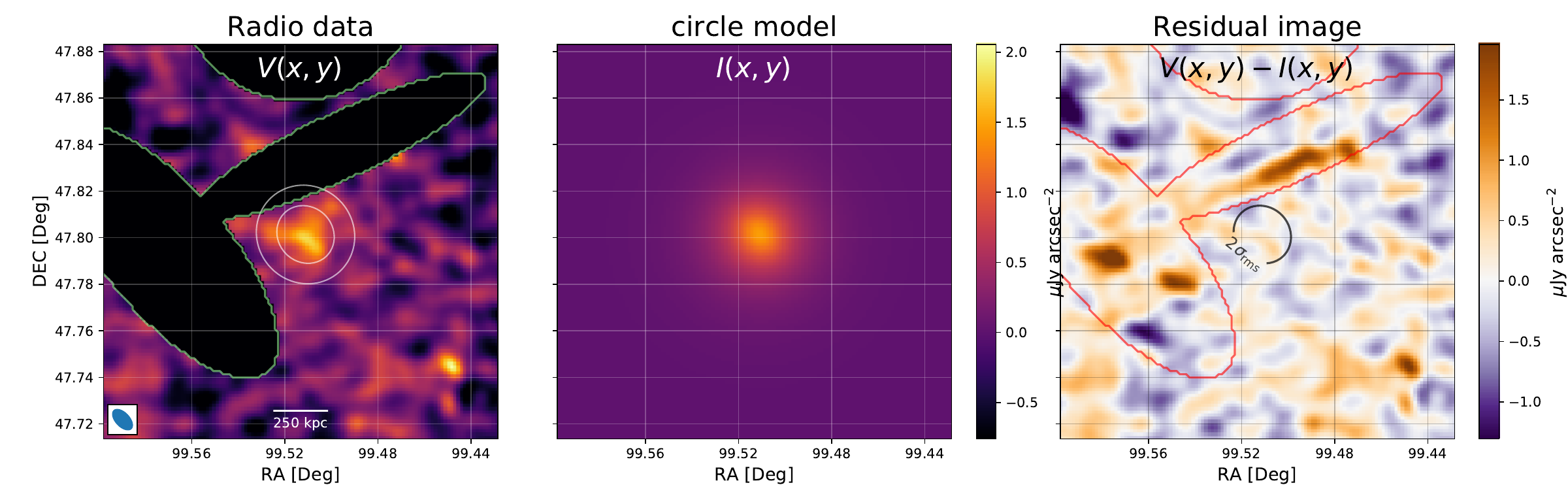}
     \end{subfigure}
     \begin{subfigure}[b]{\textwidth}
         \centering
         \includegraphics[width=\textwidth]{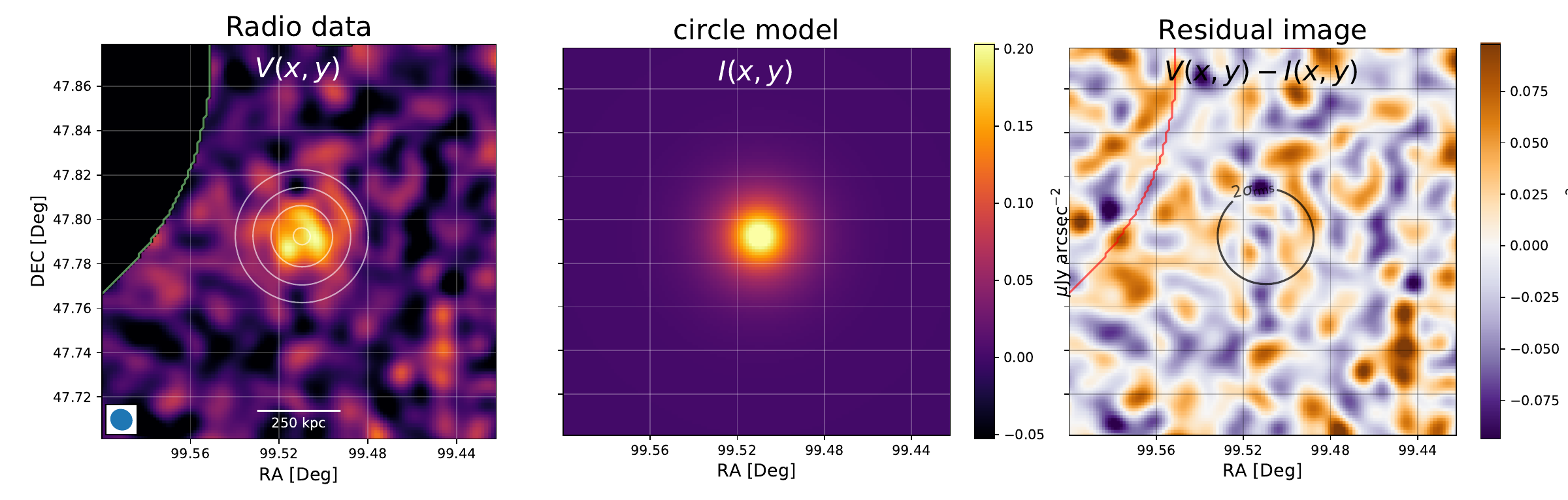}
     \end{subfigure}
     \vspace{-15pt}
     \caption{
     \texttt{FDCA-Halo} results for the radio halo in \Zwcl at LBA \textit{(first row)}, HBA \textit{(second row)}, GMRT \textit{(third row)}, \reviewfirst{VLA} \textit{(fourth row)} frequency. \textit{Left panel:} Original image with the contaminating regions masked out. \textit{Middle panel:} Circular model map. \textit{Right panel:} Residual image. The contour shows the $2\sigma_{\rm rms}$ level of the model. The red contours show the masked regions, the contamination sources are visible. 
    }
\end{figure*}

\bsp    
\label{lastpage}
\end{document}